\pdfoutput=1
\documentclass[sigconf,nonacm,authorversion]{acmart}
\usepackage{booktabs}
\usepackage{tabularx}
\usepackage{longtable}
\usepackage{needspace}
\usepackage{placeins}
\usepackage{float}
\newcommand{\PaperTableDescription}[1]{}

\newcommand{\CBase}{\textsc{Text-Reactive}}
\newcommand{\CSee}{\textsc{Visual-Reactive}}
\newcommand{\CAct}{\textsc{Visual-Proactive}}

\copyrightyear{2027}
\acmYear{2027}
\acmMonth{8}
\setcopyright{none}

\begin{document}

\title[Visual Access and Proactive AI Tutoring]{Exploring Visual Access and Proactive AI Tutoring
for Handwritten Mathematics}

\author{Nachiket Kotalwar}
\affiliation{%
  \institution{Carnegie Mellon University}
  \city{Pittsburgh}
  \state{PA}
  \country{USA}
}
\email{nkotalwa@andrew.cmu.edu}
\author{Conrad Borchers}
\affiliation{%
  \institution{Vanderbilt University}
  \city{Nashville}
  \state{TN}
  \country{USA}
}
\email{c.borchers@vanderbilt.edu}
\author{Shubham Gandhi}
\affiliation{%
  \institution{Carnegie Mellon University}
  \city{Pittsburgh}
  \state{PA}
  \country{USA}
}
\email{srgandhi@andrew.cmu.edu}

\author{Partha Sarathi Purkayastha}
\affiliation{%
  \institution{ETH Zürich}
  \city{Zürich}
  \country{Switzerland}
}
\email{ppurkayastha@ethz.ch}

\author{David Yaron}
\affiliation{%
  \institution{Carnegie Mellon University}
  \city{Pittsburgh}
  \state{PA}
  \country{USA}
}
\email{yaron@cmu.edu}

\author{Carolyn P. Ros\'{e}}
\affiliation{%
  \institution{Carnegie Mellon University}
  \city{Pittsburgh}
  \state{PA}
  \country{USA}
}
\email{cprose@cs.cmu.edu}

\renewcommand{\shortauthors}{Kotalwar et al.}

\begin{abstract}
\looseness=-1
Handwritten work captures mathematical reasoning that learners may not express in typed or voice-dictated messages. We built a multimodal AI tutoring platform in which a phone camera gives the tutor a live view of learners' work on paper. In a randomized online study, 117 adults solved calculus problems with \CBase{}, \CSee{}, or \CAct{} assistance. Analysis of 2,140 tutor messages showed that visual access shifted assistance toward checking ongoing work, while proactivity enhanced error identification and directive hints. Learners valued feedback grounded in their writing; those who received proactive help rated it as supportive, and about a third wanted it more often. These findings inform when tutors should intervene and what they should address. We release the platform, prompts, and instruments, together with 255,984 timestamped images of handwritten work linked to tutoring conversations, intervention decisions, and learning outcomes, to support multimodal tutoring systems that preserve the affordances of handwritten work while providing adaptive feedback.
\end{abstract}



\maketitle
\hypersetup{pdfsubject={Human-centered computing: Empirical studies in HCI; Applied computing: Education},pdfdisplaydoctitle=true}

\section{Introduction}

Effective tutoring adapts to learner problem-solving behavior. A human tutor follows a
learner's solution as it develops, notices an error at the step where it
occurs, and decides moment by moment whether to intervene or let the
learner continue~\cite{merrill1992effective, chi2001learning,
vandepol2010scaffolding}. VanLehn~\cite{vanlehn2011relative} found that computer-based tutoring with feedback and scaffolding at the level of individual steps was nearly as effective as human tutoring and more effective than answer-level feedback. Meta-analyses show that intelligent tutoring systems built on this principle improve learning outcomes~\cite{kulik2016effectiveness, ma2014intelligent, steenbergenhu2014college}. Such step-level support requires that the tutor can see what the learner is doing.

Step-based tutoring systems can follow intermediate steps because learners enter them through the system's interface. Much mathematical work, however, happens on paper, through diagrams, symbolic expressions, and revised intermediate steps. Handwriting input lets learners record equations faster and with fewer errors than typing~\cite{anthony2012paradigm}, and entering work through a keyboard changes how learners work through problems~\cite{demorais2022handwriting}. Borchers et al.~\cite{borchers2023paper} call for combining the flexibility of paper practice with the support of tutoring systems. Chat-based AI tutors see only what learners describe or separately share, and the learners who most need help are often the least able to recognize and
articulate that need~\cite{aleven2003helpseeking}. How can an AI tutor engage with
this work without requiring learners to reproduce it in another
interface?

\begin{figure*}[t]
  \centering
  \includegraphics[width=\linewidth,alt={Diagram of the paper-and-phone tutoring setup and three assistance conditions. A supported phone records handwritten work while a laptop presents the problem and tutor conversation. Problem context, conversation, and tutoring guidelines are shared. Text-Reactive answers learner requests without seeing the paper. Visual-Reactive also receives the current paper view when answering. Visual-Proactive additionally observes between requests and can initiate help. Arrows distinguish learner requests, tutor responses, paper access, and proactive assistance.}]{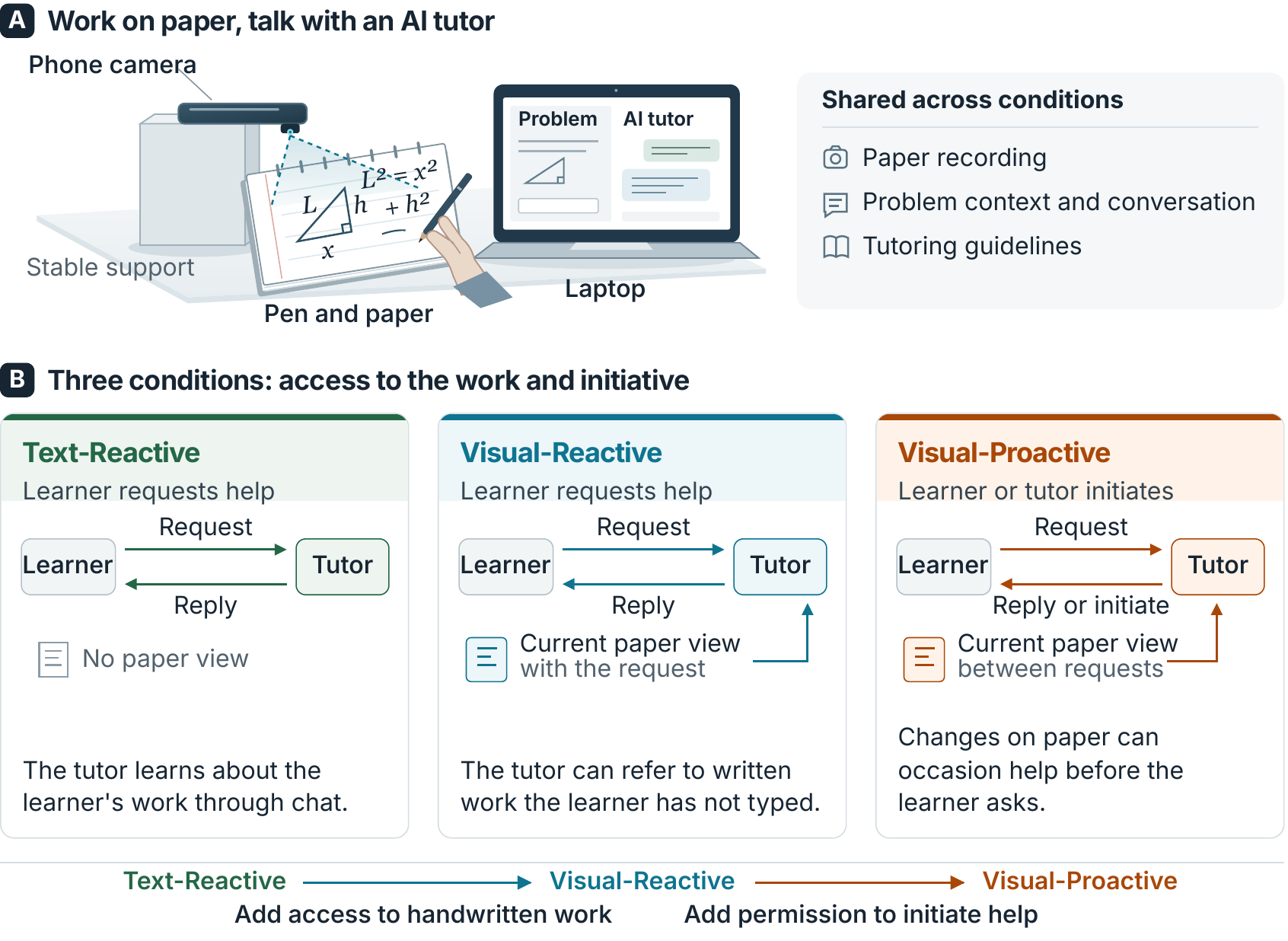}
  \caption{Paper-based tutoring and the three assistance conditions.
  \emph{(A)} A supported phone records the learner's paper while a laptop
  presents problems and tutor chat. \emph{(B)} In \CBase{} the tutor learns about the
  learner's work only through chat; \CSee{} can additionally use the current
  paper view when responding to a request; \CAct{} can also consult the
  view between requests and initiate help. The bottom arrows identify
  the capabilities added between neighboring conditions. Each learner
  participates in one condition.}
  \Description{Panel A shows a phone on a stable support overlooking a
  notebook, a hand writing with a pen, and a laptop with problem and chat
  panes. Paper recording, problem context and conversation, and
  tutoring guidelines are common to all conditions. Panel B shows three
  side-by-side conditions, each with learner-to-tutor request and
  tutor-to-learner response arrows. Text-Reactive has no tutor access to
  the paper view. Visual-Reactive adds a paper-view input when answering
  a request. Visual-Proactive also uses the view between requests and
  can initiate help. Two named condition-to-condition arrows identify
  the additions of paper access and permission to initiate help.}
  \label{fig:system}
\end{figure*}
Research on collaborative physical tasks shows what a shared view of the
work supplies. When a helper can see the worker's task space, pairs
need fewer words to establish reference, actions can stand in for
spoken confirmations, and the helper can monitor progress and offer
help at the right moment~\cite{fussell2000sharedvisual, kraut2003visual,
gergle2004action}. The view serves both grounding and awareness of the
task state~\cite{gergle2013grounding}. In stepwise mathematical problem solving, a useful hint depends on the step the learner is on now. When learners work on paper, a view of the page gives the tutor access to intermediate expressions and revisions that may never appear in the conversation.

Earlier systems connected handwritten work to tutoring through instrumented pens, tablets, or interactive whiteboards~\cite{anthony2012paradigm, lee2007newtonspen, demorais2022handwriting, lee2025sketchpad, liu2026amiwrite}, which require dedicated hardware and move writing away from the learner's usual paper. Current multimodal AI is increasingly able to read handwritten expressions
and diagrams from camera images of paper work and reason about them in
conversation~\cite{baral2025drawedumath, nath2025fermat,
yang2026mmtutorbench}. This makes it practical to support work on ordinary paper with a phone and laptop that learners already own. We built a platform in
which a phone camera, supported above ordinary paper, gives a multimodal
AI tutor a live view of the learner's handwritten work while a laptop
presents problems and chat (Figure~\ref{fig:system}A).  Establishing a stable, readable view is a practical
demand on the learner, and we report how participants managed it.

A live view can change help in two ways. When a learner asks for help,
the tutor can read the written step instead of requiring the learner to
retype an expression or describe a diagram~\cite{clark1991grounding}.
Between requests, the view supports awareness of the developing work,
providing a basis for \emph{proactivity}: initiating help in response to
work the learner has not described. Reactive help depends on the learner recognizing a difficulty and formulating a request~\cite{aleven2003helpseeking}. Proactive help can reach a learner before a request, but may interrupt productive work or supply a step the learner could have developed independently~\cite{horvitz1999mixedinitiative, koedinger2007assistance}. Evidence on this trade-off for AI tutoring is still limited: in video-based learning, Otmani et al.~\cite{otmani2026proactive} found that proactive mentoring increased interaction while reactive mentoring left learners with greater perceived choice and control, and knowledge gains did not differ significantly. A tutor that observes handwritten work adds a further possibility, initiating help in response to the developing solution rather than on a schedule.

We therefore compare three forms of assistance (Figure~\ref{fig:system}B). \CBase{} responds to learner messages without seeing the paper; \CSee{} also sees the current paper view when answering a message; \CAct{} additionally observes the work between requests and can initiate help. Comparing \CBase{} with \CSee{} asks what visual access adds when the learner asks for help; comparing \CSee{} with \CAct{} asks what ongoing observation and tutor initiative add when the view is already available. For both contrasts we examine learning, the assistance learners received, and how they experienced it.

We studied these conditions online with 117 adults who solved
introductory calculus problems in an assisted activity bracketed by
unaided pre- and post-tests. We address three research questions:
\begin{itemize}
  \item \textbf{RQ1.} How do tutor use, problems attempted during the
  assisted activity, and pre-to-post learning gains vary across the
  three assistance conditions?
  \item \textbf{RQ2.} How do visual access and proactivity shape learners'
  help seeking and the assistance provided by the tutor?
  \item \textbf{RQ3.} How do learners experience the tutoring assistance
  and the paper-and-phone setup?
\end{itemize}

Learners in all three conditions improved significantly from pre-test to post-test. Adjusted post-test scores did not differ detectably between conditions, and an exploratory analysis pointed to a possible advantage of proactive tutoring for learners with low prior knowledge (Section~\ref{sec:learning-results}). Visual access
shifted the tutor's messages toward checking ongoing work, and requests
leaned the same way; proactive messages added error identification and directive hints.
Learners who received proactive help rated it as supportive rather than intrusive; learners in both visual conditions rated the help as more useful because they did not need to restate their written steps; and a recurring wish across conditions was for more initiative from the
tutor, although learners differed in how much guidance they wanted. Because more assistance is not always what best supports learning, these preferences are best judged alongside learning outcomes~\cite{koedinger2007assistance}. Together, these results show that a tutor with a live view of ordinary
paper work is feasible with everyday devices, and that the view shifts
assistance toward the learner's actual work. They also raise the question
of how an AI tutor that sees the work should choose the timing, purpose,
and depth of its help. We release the platform source, tutoring prompts, instruments, and an interaction dataset of 255,984
timestamped images of handwritten work, covering 66.7 hours of assisted
activity, aligned with conversations, intervention decisions, and
learning outcomes (Section~\ref{sec:dataset}). These resources support
further research on these questions.

\section{Related Work}

\subsection{Seeing the Work: Shared Visual Context in Tutoring and Collaboration}

Studies of human tutors show how much of their effectiveness rests on
watching the work. Merrill et al.~\cite{merrill1992effective} concluded that human tutors
let students do as much of the work as possible while intervening on
errors quickly enough to prevent floundering. Compared with the
tutoring systems of the time, they left more of the error recovery to
the student.
VanLehn et al.~\cite{vanlehn2003impasse} found that learning during tutoring concentrated at impasses, the moments when a
student got stuck, made an error, or acted with uncertainty. Noticing an
impasse depends on observing the work as it develops.
Scaffolding research reaches the same conclusion from the
tutor's side: contingent support requires continuous diagnosis of what
the learner has just done~\cite{wood1976tutoring, vandepol2011contingent}.

Research on shared visual space in collaboration shows what a view of
the work supplies and what it costs. Fussell et al.~\cite{fussell2000sharedvisual} found that pairs repairing
a bicycle side by side worked about a quarter faster than remote pairs
and referred to parts with far fewer words, but a head-mounted video
link did not recover that advantage. In a follow-up, a view of the
workspace let remote helpers see when the worker was stuck or ready for
the next step and time their instructions accordingly, although keeping
the camera on the relevant part took effort of its own~\cite{kraut2003visual}. Gergle
et al.~\cite{gergle2013grounding} distinguished two functions of a shared
view, grounding what is said and maintaining awareness of the task state,
and showed that delay and a restricted field of view reduced these
benefits. Visible actions could also replace spoken
confirmations~\cite{gergle2004action}.

These findings come from human pairs. Whether an AI system can use a
live view in the same way is a newer question. Bohus et al.~\cite{bohus2024isthisit} observed people working through
physical tasks with a head-worn multimodal assistant and found that most
of their questions sought to ground its instructions in the objects,
states, and actions in front of them; they identified tracking the user's
actions and deciding when to intervene as open problems. Shaikh et al.~\cite{shaikh2024groundinggaps}
showed that language models perform the grounding acts human
interlocutors use, such as clarification and acknowledgment, far less
often than people do. In mathematics, vision-language models underperformed on struggling
students' handwritten work and misdiagnosed their errors, even when the
images were redrawn cleanly~\cite{baral2025drawedumath, lucy2026aftermath, nath2025fermat}.
Learners, in turn, must judge whether the model has read their work
correctly; Nguyen et al.~\cite{nguyen2024misread} described how novices
and code models misread each other and how many attempts novices spent
repairing prompts when their intent had been misread.

These studies establish that a shared view helps human helpers
and that current models can use one, imperfectly. Our \CBase{} and
\CSee{} conditions isolate what a view of ordinary paper adds while
learners solve problems, and Section~\ref{sec:process-results} reports
how it changed what learners asked and what the tutor did.

\subsection{Initiating Help: Proactive Assistance and Learner Control}
\label{sec:related-proactive}

Whether a tutor should wait to be asked is an old question in learning
environments, because help seeking is itself demanding. A review by Aleven et al.~\cite{aleven2003helpseeking} concluded that
students often fail to seek help when they need it, or use help in ways
that avoid thinking, such as requesting hints until the answer
appears~\cite{baker2004gaming}. Tutoring systems have therefore tried to detect need
from behavior, predicting unproductive struggle from interaction
logs~\cite{maniktala2022helpneed} or disengagement from
gaze~\cite{dmello2012gazetutor}, and to teach better help
seeking~\cite{roll2011helpseeking}. Generative AI makes asking more
demanding still: learners must know what to ask and how to judge what
comes back~\cite{tankelevitch2024metacognitive}, and in a semester-long
deployment of a guard-railed programming assistant, requests ranged from
conceptual questions to attempts to obtain the
solution~\cite{kazemitabaar2024codeaid}. Intervening too readily has its
own cost. The assistance dilemma~\cite{koedinger2007assistance} frames
intervention as a balance between giving and withholding help, each with
costs, and research on productive failure~\cite{kapur2008productive} suggested that support arriving before learners have struggled can remove
the opportunity that produces learning.

Mixed-initiative research frames the decision to intervene as one made
under uncertainty about the user's goals and the cost of a poorly timed
interruption~\cite{horvitz1999mixedinitiative}. Interruptions are less
disruptive at task boundaries~\cite{iqbal2008defer,
adamczyk2004breakpoints}, and whether they arrive immediately, at a
negotiated moment, or on a schedule changes their
cost~\cite{mcfarlane2002coordinating}. Recent proactive assistants built on large language models (LLMs)
apply these ideas by observing ongoing work and deciding when to speak.
In programming, Codellaborator's inactivity-triggered suggestions were
sometimes experienced as interruptions because they arrived while
participants were still thinking~\cite{pu2025codellaborator}. Chen et al.~\cite{chen2025needhelp}
found that an assistant that chose when to interject from the editor
state improved productivity over a reactive one. Participants preferred
that assistant only when its interjections were spaced out and limited. Proactive assistants outside programming decide when to intervene from
what they hear or see, such as the user's
speech~\cite{tabalba2024articulatepro} or a head-worn
camera~\cite{li2025satori, bandyopadhyay2025yeti, pu2025promemassist},
and recent work adapts the
initiation policy to individual preferences~\cite{wang2026initiation}.
In learning, proactive mentoring around an instructional video increased interaction~\cite{otmani2026proactive}, and Li
et al.~\cite{li2026timing} examined the tension between timely
intervention by process-aware tutors and learners' boundaries on what
may be observed.

These systems observe code, speech, or camera streams, mostly in short
tasks with small samples, and only Otmani et al.~\cite{otmani2026proactive} measured learning.
Our \CAct{} condition adds initiative
to a tutor that can see the paper. We report what it chose to address (Sections~\ref{sec:proactive-results}--\ref{sec:proactive-decisions}), whether learners found its
messages interrupting (Section~\ref{sec:experience-results}), and
learning outcomes alongside
both (Section~\ref{sec:learning-results}).

\subsection{AI Tutoring for Mathematics with Handwritten Work}
\label{sec:related-tutoring}

This work builds on the tradition of step-level tutoring. VanLehn's ``inner loop'' gives feedback and hints on each step within a problem~\cite{vanlehn2006behavior}, and tutorial dialogue systems build explanations through
conversation~\cite{graesser2001conversational}. Both require the
learner's steps to be visible to the system, so intelligent
tutors have long required work to be entered through structured
interfaces~\cite{gertner2000andes}. LLM-based tutors relax the interface
but keep the requirement. Dialogue corpora and tutoring benchmarks are organized around typed
student turns~\cite{macina2023mathdial,
macina2025mathtutorbench}, and efforts to align model responses with
pedagogy address what the tutor says, not what it can
see~\cite{sonkar2024pedagogical, daheim2024stepwise, maurya2025unifying}.
Controlled evaluations of LLM tutoring report mixed outcomes. AI tutoring outperformed in-class active learning in a physics
course~\cite{kestin2025aitutoring}. In a UK classroom trial, a
pedagogically tuned model supervised by human tutors supported learning
with no detected difference from the tutors themselves~\cite{learnlm2025rct}.
Mathematics studies found gains from model-generated help comparable to
human-authored hints~\cite{pardos2024chatgpt} and from a messaging-based
tutor in Ghanaian schools~\cite{henkel2024ghana}. Unrestricted access to
a generative model, however, harmed high-school mathematics
learning~\cite{bastani2025guardrails}, and in programming, novices who were already succeeding used generative
AI to accelerate while struggling novices were left with an illusion of
competence~\cite{prather2024wideninggap}. Studies of
AI-supported learning also distinguish assisted performance from later
unaided performance~\cite{kazemitabaar2023codegenerators}, a distinction
our design adopts.

The typed-input requirement carries a particular cost in mathematics.
Students expressed and solved problems differently by hand than by
keyboard~\cite{anthony2012paradigm, anthony2007benefits}. Systems that tutor handwritten work have therefore built dedicated input surfaces~\cite{lee2007newtonspen, demorais2022handwriting, liu2026amiwrite, lee2025sketchpad}. Photo-based math apps go the other way: they read a single picture of a problem, and submission-speed data suggest students increasingly use them to get
answers quickly~\cite{sloanlynch2022photoapps}. Vision-language models can now read handwritten
mathematics~\cite{gervais2025mathwriting, nath2025fermat}, including
phone photographs of paper work~\cite{baral2025drawedumath}, and reason over handwritten
solution steps in tutoring settings~\cite{yang2026mmtutorbench}, though
their error rates remain substantial. This makes a view of ordinary paper a plausible, if imperfect, input for a tutor.

We are not aware of work testing what a view of ordinary paper does for
tutoring. Studies of handwriting input compared input modalities rather than what the tutor could see~\cite{anthony2007benefits, demorais2022handwriting}. Handwriting tutors are described as systems or
compared on the format of their feedback, without a condition that lacks
the view and rarely with learning outcomes. LLM tutoring trials measure
learning, but with typed input. We compare a
tutor with and without a view of ordinary paper and, with the view, with and without initiative, among 117 adults, and report learning gains alongside a
content analysis of what learners asked and what the tutor did
(Section~\ref{sec:results}).

\section{Study Platform and Assistance Conditions}
\label{sec:system}

We designed a common paper-and-phone platform to compare visual access
and proactive assistance while keeping the interface, task materials,
and tutoring policy consistent across conditions
(Figure~\ref{fig:system}).

\subsection{Platform Setup}

\begin{figure*}[!htbp]
\centering
\includegraphics[width=\linewidth,alt={Phone and laptop screenshots of the tutoring platform. The phone camera shows handwritten work for a changing-rectangle problem. The laptop places the problem, numeric answer field, and timer beside the tutor conversation, with an expandable preview of the same paper. Learners can type or dictate messages. The paper view and the three chat messages come from one participant's session: a proactive tutor message about the written answer, the learner's question, and the tutor's reply. The header, timer, and status text are illustrative.}]{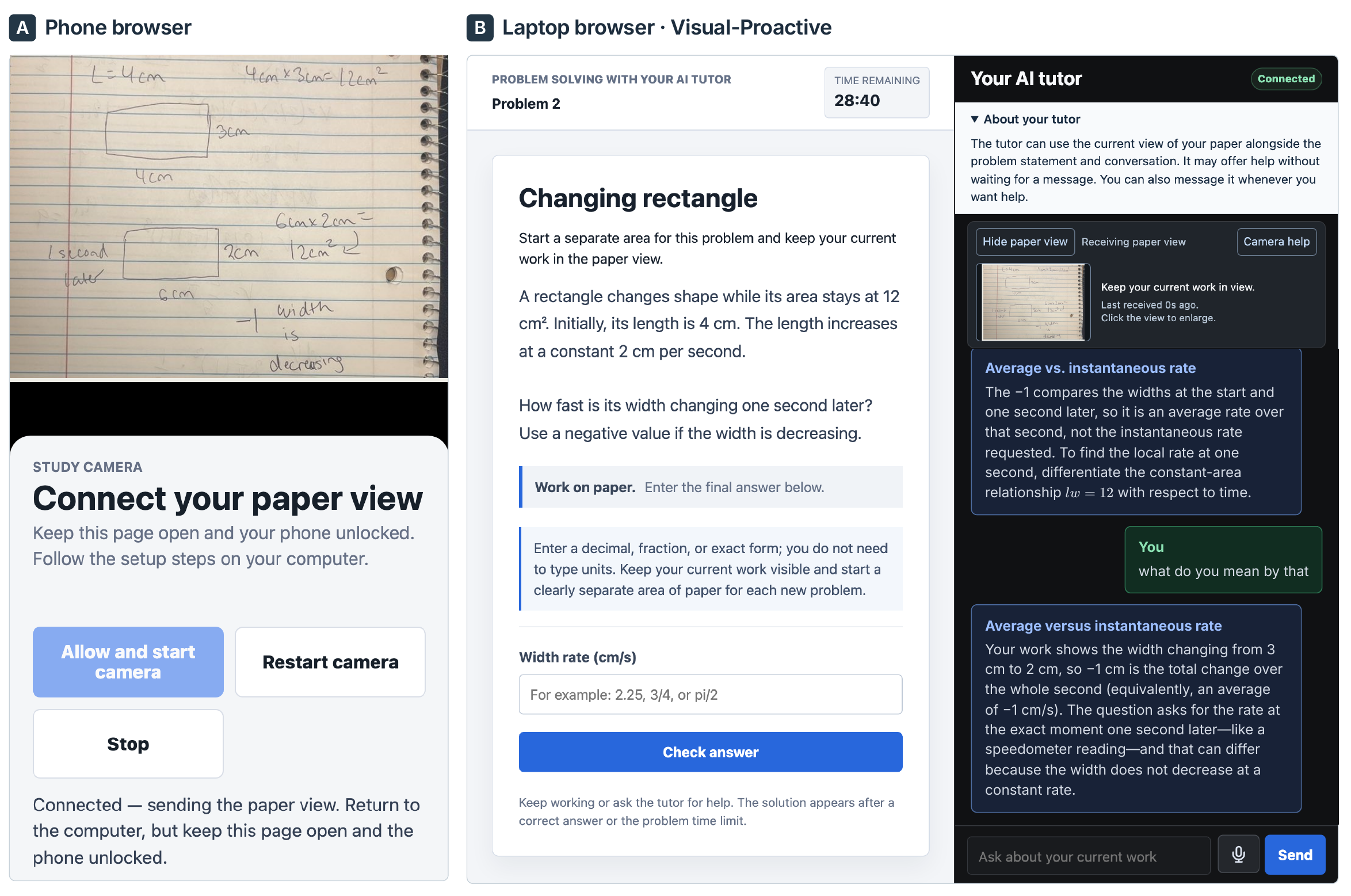}
\caption{The browser-based study platform, shown with one participant's
paper view and exchange from the study. (A) Scanning the laptop's
QR code opens the phone camera interface without an app installation.
(B) The laptop presents the problem and answer entry beside the tutor
conversation. The \CAct{} interface is shown with tutor information visible and the
optional paper preview expanded. The first tutor message was sent
proactively after the learner wrote $-1$; the learner then asked a
follow-up question. The problem header, timer, and status text are illustrative.}
\Description{Two screenshots show a connected phone camera page and the
laptop tutoring interface. The phone displays a participant's work on the changing rectangle problem. The same work appears in the laptop's paper preview,
above a proactive tutor message, the learner's question, and the tutor's reply. The laptop also shows the
problem, answer field, timer, text entry, and voice-dictation control.}
\label{fig:platform-interface}
\end{figure*}

We developed a guided browser-based setup so learners could establish a
readable paper view with their own devices, without installing an app.
Learners scan a QR code on the laptop with their phone, which opens the phone's camera page.
Learners allow camera access and support the phone above their paper,
leaving both hands available for writing. A laptop preview helps them
check the view. Figure~\ref{fig:platform-interface} shows the interfaces;
Appendix~\ref{app:setup-screens} illustrates setup; the \texttt{study-materials/} folder of the project repository (Appendix~\ref{app:tutor-materials}) provides setup and tutorial instructions. We report exits during camera setup (Section~\ref{sec:participants})
and participants' ratings of setup ease
(Section~\ref{sec:experience-results}).

\subsection{Solving Problems with the Tutor}

The laptop presents the problem, a final-answer field, and tutor chat. Learners work on paper and request help by typing
or voice dictation. After a worked solution becomes available, they can
review it with the tutor before proceeding to the next problem, which
starts a new conversation. Section~\ref{sec:procedure} gives the timing
and worked-solution release rules.

\subsection{Assistance Conditions}
\label{sec:conditions}

We used GPT-5.6 Sol as a multimodal AI tutor. All conditions record paper
views and share task materials, the chat interface, and tutoring policy.
The tutor receives the current problem, a private reference solution,
the problem's conversation, and whether the learner is still attempting
the problem or reviewing its worked solution.
Its access to paper views and permission to initiate help vary:

\begin{itemize}
  \item \CBase{} responds to learner messages using the problem context
  and conversation, without access to the paper view.
  \item \CSee{} additionally receives the current paper view with each
  learner message, allowing requests to refer to work not entered in
  chat. It responds only to learner messages.
  \item \CAct{} uses the same requested-help policy as \CSee{}, but also
  observes paper views over time and can initiate help based on ongoing
  work.
\end{itemize}

Throughout the paper, \emph{requested replies} are the tutor's answers to
learner messages, and \emph{proactive messages} are messages the tutor
initiates. Only \CAct{} sends proactive messages.
The two pairwise contrasts between neighboring conditions add visual access during requested help,
then observation and initiative between requests (Figure~\ref{fig:system}B).
We omitted a
text-proactive condition because timed check-ins without visual access
could not respond to changes in written work, the basis for proactivity
here.

\subsection{Tutoring Policies}
\label{sec:tutoring-policies}

\paragraph{Shared pedagogical policy.}
System prompts encode a shared pedagogical policy, condition-specific
requested-help instructions, and a proactive-decision policy. The application controls which conditions receive paper views and when
the worked solution is released.
Appendix~\ref{app:tutor-materials} gives further detail and points to
the complete prompts and source code in the project repository.

The policy draws on step-level tutoring~\cite{vanlehn2006behavior}, contingent scaffolding~\cite{vandepol2010scaffolding}, and explanatory feedback~\cite{shute2008formative}. It instructs the tutor to
build on the learner's own approach, notation, and step order whenever
these are mathematically valid, explain the reasoning behind suggested
steps, and leave meaningful work to the learner. The reference solution
shows one correct path, and the tutor may depart from it.

Learners can request help at any point. To balance assistance and learner
effort~\cite{koedinger2007assistance}, when a learner repeatedly asks
for the next step without trying it, the tutor is instructed to help them
attempt the current step instead. Learners expressing difficulty or requesting a demonstration can
receive a worked intermediate step, leaving the next step to them.
Evidence that a learner has carried out a step can come from the paper
view, from chat, or from the learner's own report, depending on
condition; the tutor does not require learners to type out their work or
confirm each step.

During attempts, the tutor is instructed to withhold the final answer
and, beyond such a worked intermediate step, avoid completing steps for the learner. It can check proposed answers and help
repair incorrect reasoning. Once the worked solution is revealed, the
tutor can explain the solution in full. These boundaries apply in every condition. The
reference solution supports the tutor's correctness checks but does not
guarantee that they are right, so we checked the mathematical correctness of a sample of tutor
messages (Appendix~\ref{app:math-correctness}).

\paragraph{Proactive intervention.}
A tutor that can see the work can intervene contingently, and must then
decide when an observation calls for a message. The
\CAct{} design addresses that decision directly. At regular intervals during an attempt, the tutor receives the latest paper view
together with earlier views and the history of delivered help, and
chooses one of three actions: repair a clear, consequential error;
support a transition when a valid partial approach has stalled at the
same unresolved step; or wait. Waiting is the default. The tutor waits when the learner's writing is
still progressing between views, when it has just given help that the
learner has not yet had time to use, or when the page is blank or does
not show the current problem. A pause alone is not treated as evidence
that a learner is stuck~\cite{horvitz1999mixedinitiative}.

The two intervening actions differ in the evidence they require and in
what they deliver. A repair needs a single observation of a consequential
error. It identifies the fault, explains why it matters, and gives one hint for
repairing it without supplying the corrected expression. Transition
support needs persistence: the same unresolved operation across comparable views for a waiting interval, timed from the last meaningful progress or the last delivered help, whichever is more recent. Because
a repair can follow a single observation while transition support must
wait for persistence, we call this the repair-first policy. Proactive messages
stay within the same boundaries on mathematical assistance as requested
help.

A quiet period after every learner or tutor message keeps proactive help
from crowding the conversation, and learner requests take priority.
Proactive messages stop once the worked solution is shown. A brief
chime draws attention to proactive messages; learners can keep writing or
respond in chat, and the tutor is instructed to acknowledge requests to
work independently. Appendix~\ref{app:tutor-materials} gives the analysis
interval, quiet period, waiting interval, and per-session limit;
Appendix~\ref{app:platform-refinement} describes how we refined the paper
view and the delivery of proactive messages after pilot sessions; and
Section~\ref{sec:proactive-decisions} reports how often the tutor chose
each action.

\section{Study Method}

\subsection{Participants and Recruitment}
\label{sec:participants}

We recruited adults with prior calculus experience to examine support
for previously studied concepts and the practicality of phone-based
tutoring in their own workspaces. Recruitment was on Prolific, among
adults aged 18 or older in the United States who had studied differential calculus through
coursework or structured self-study and passed a brief screening.
Participants needed a laptop or desktop computer, a camera-equipped
smartphone, and pen and paper. Compensation was US\$20 per hour.

Prospective participants first saw the paper-and-phone setup requirements; 74 chose not to proceed at that point, for reasons we did not record. Of the 169 people who entered the study, 117 completed it. The 52 who exited did so during camera setup, most citing problems getting the phone camera connected and started (25) or difficulty keeping a stable, readable view while writing (19). The setup was identical in every condition and
preceded any condition-specific experience.
Participants were assigned to conditions by simple randomization: 40 in
\CBase{}, 37 in \CSee{}, and 40 in \CAct{}. Participants also reported their preferred working
medium, how recently they had used calculus, their AI use for
mathematics, and the courses they had taken
(Table~\ref{tab:participant-background});
Appendix~\ref{app:instruments} gives screening criteria.

\begin{table*}[!htbp]
\centering
\small
\caption{Self-reported working habits and prior experience. Entries are participant counts. Single-choice response categories were combined for presentation: digital tools include digital handwriting, typed documents, and mathematics software; other/no clear preference includes mental calculation; occasional AI use means monthly or less often. Recency within the past year includes current use. Completed courses allow multiple
selections.}
\label{tab:participant-background}
\begin{tabularx}{\linewidth}{Xrrrr}
\toprule
Background & \CBase{} & \CSee{} & \CAct{} & Overall \\
$n$ & 40 & 37 & 40 & 117 \\
\midrule
\addlinespace[3pt]
\multicolumn{5}{l}{\textit{Preferred working medium}} \\
\quad Paper & 22 & 20 & 25 & 67 \\
\quad Mixed physical and digital & 13 & 9 & 7 & 29 \\
\quad Digital tools & 3 & 3 & 5 & 11 \\
\quad Other / no clear preference & 2 & 5 & 3 & 10 \\
\addlinespace[3pt]
\multicolumn{5}{l}{\textit{Most recent use of calculus (derivatives)}} \\
\quad Within the past year & 9 & 11 & 11 & 31 \\
\quad 1--3 years ago & 10 & 7 & 11 & 28 \\
\quad More than 3 years ago & 21 & 19 & 18 & 58 \\
\addlinespace[3pt]
\multicolumn{5}{l}{\textit{AI use for mathematics in the past year}} \\
\quad Never & 16 & 7 & 11 & 34 \\
\quad Occasionally & 15 & 23 & 17 & 55 \\
\quad At least weekly & 9 & 7 & 12 & 28 \\
\addlinespace[3pt]
\multicolumn{5}{l}{\textit{Completed mathematics courses (multiple selections)}} \\
\quad Algebra II or equivalent & 38 & 33 & 33 & 104 \\
\quad Precalculus or trigonometry & 37 & 31 & 29 & 97 \\
\quad Calculus I or equivalent & 35 & 35 & 37 & 107 \\
\quad Calculus II or equivalent & 25 & 24 & 25 & 74 \\
\quad Multivariable calculus & 18 & 10 & 9 & 37 \\
\quad Differential equations & 19 & 15 & 10 & 44 \\
\quad Other & 2 & 1 & 0 & 3 \\
\bottomrule
\end{tabularx}
\end{table*}

The study was approved by our institution's Institutional Review Board.
Participants provided electronic informed consent and could withdraw at
any time. The consent materials described the collection of paper images,
conversations, and survey recordings, including processing by third-party
AI services. Participants were instructed to direct the camera toward
their paper and keep personal information out of the view and chat.
Study records were stored on university-controlled systems with access
restricted to the research team.

\subsection{Tasks and Tests}

The assisted activity used related-rates problems from introductory
differential calculus. These problems ask how fast one quantity changes
given the rate of another, and solving them means setting up a
relationship between changing quantities, differentiating it, and
substituting values. Learners often do this work on paper.

All conditions used the same ordered bank of ten problems and worked
solutions. Early problems covered the concepts the tests assess: rates under
constraints, geometric relationships and the chain rule, and how a rate
depends on the current state. Later problems provided further
practice and combined concepts. This ordering gave learners a common
starting point while letting faster learners reach additional problems
within the activity period. Each required one numeric final answer.
Appendix~\ref{app:curriculum-sequence} lists the sequence and conceptual focus.

The pre-test and post-test used alternate nine-item forms covering the
same concepts through different questions, completed without AI
assistance, with form order counterbalanced within each condition. Pilot sessions with the same recruitment population helped us calibrate
wording, difficulty, scoring, timing, and consistency between forms (Appendix~\ref{app:instruments}).
The repository's \texttt{study-materials/} folder includes both test
forms with scoring keys and all assisted problems with worked solutions.

We estimated internal consistency for the 117 participants using
Kuder--Richardson Formula~20 (KR-20)~\cite{kuder1937reliability}, separately
for each form at pre-test and post-test because consistency can vary
across administrations~\cite{edelsbrunner2025alpha}. Estimates ranged
from .77 to .85; Table~\ref{tab:test-reliability} in
Appendix~\ref{app:instruments} reports all four estimates and their 95\%
confidence intervals.

\subsection{Study Design and Procedure}
\label{sec:procedure}

Each participant completed one session of about 80 minutes: a
background questionnaire, a 12-minute pre-test, camera setup, a 35-minute assisted activity with the tutor, a
12-minute post-test, and a final survey.

After the pre-test, participants followed standardized camera-setup
instructions. The common camera arrangement kept the physical setup
comparable and recorded written work in every condition. An on-screen introduction explained the activity and what the assigned
tutor could do, including visual access and proactive help in the
conditions that had them.

During the 35-minute assisted activity, participants worked through as
many problems as they could in sequence. Worked solutions became
available after a correct answer or ten minutes on a problem, regardless
of intermediate progress. Learners could review each solution with the
tutor before continuing; solving and review shared the 35-minute period.
The session ended with the 12-minute unaided post-test and final survey.
Conditions therefore shared time limits, ordered materials, and
worked-solution release rules.
Section~\ref{sec:completion-results} reports problems attempted.

The survey used individual seven-point agreement items about setup ease,
communication, feedback fit, and control over the next mathematical step,
with additional items for visual and proactive capabilities. Participants
could select ``Not applicable /
Could not judge.'' Two open questions and an optional comment followed
(Section~\ref{sec:thematic-method}). We asked participants to speak their answers into a recording and
offered typing as an alternative; we transcribed the recordings automatically. Table~\ref{tab:survey-ratings} lists the rated items, and Appendix~\ref{app:instruments} describes their administration; the repository's \texttt{study-materials/} folder provides
the full background and final questionnaires.

We recorded test and survey responses, timestamped paper views,
conversations, answer submissions, problem transitions, worked-solution
exposure, and, in \CAct{}, each of the tutor's paper analyses and
intervention decisions. Inference cost averaged US\$1.93 per session,
higher in \CAct{} because of its periodic paper analyses.

\subsection{Data Analysis}
\label{sec:data-analysis}

Tutor use, problems attempted, and learning gains addressed RQ1;
content analysis addressed RQ2; survey ratings and thematic analysis of
participant accounts addressed RQ3.

\subsubsection{Tutor Use, Problems Attempted, and Learning Gains}
\label{sec:quantitative-method}

We chose a statistical test for each measure according to its
distribution. Test scores were analyzed with linear models adjusted for
pre-test score. Message counts, problems attempted,
and single survey items, which are skewed or ordinal, were compared
across conditions with Kruskal--Wallis tests, followed by pairwise
Mann--Whitney $U$ tests~\cite{kaptein2010likert}. Pairwise comparisons
within a measure, and the label tests within each label set of the
content analysis, are Holm-corrected. We report effect sizes with the main tests; each table names the measure it uses.

\paragraph{Tutor use.}
To describe how much learners interacted with the tutor and who
initiated help, we counted learner messages and tutor messages; in
\CAct{}, we distinguished requested replies from proactive messages. We
summarized counts as means, standard deviations, and medians for the
whole activity and separately for problem solving and worked-solution
review. Messages before the worked solution was shown counted toward
problem solving; messages afterward counted toward worked-solution
review.

\paragraph{Proactive decisions.}
In \CAct{}, we counted the tutor's paper analyses, its decisions to wait
or intervene, and the messages it delivered.

\paragraph{Problems attempted.}
We counted the distinct problems each participant opened, which we call
problems attempted. We also checked whether condition differences
depended on prior knowledge, using the exploratory prior-knowledge model
described under learning gains below.

\paragraph{Learning gains.}
Both tests were scored out of nine, with one point per correct item.
Unsubmitted or ``I don't know'' responses scored zero. We tested pre-to-post change on the whole sample with a two-sided paired
$t$-test, with Cohen's $d$ computed as the mean gain divided by the
standard deviation of participants' score changes.

To compare conditions, an analysis of covariance (ANCOVA) adjusted each
participant's post-test score for their pre-test
score~\cite{vickers2001baseline,petscher2011adjusted}, with robust
standard errors; slopes were not statistically distinguishable across conditions (OLS interaction $p=.201$; Section~\ref{sec:learning-results}), and we report adjusted differences at the covariate mean, where they do not depend on the slope assumption. Residuals were approximately normal with equal variances. We did not detect an effect of test-form order on gains
($t(115)=-0.52$, $p=.602$). Scores are bounded at zero and nine: 14 participants scored nine and 6 scored zero on the post-test, and 20 scored zero on the pre-test, so fitted values near the bounds are approximations. A rank-based version of the condition model gave the same result ($p=.324$). We report adjusted means, pairwise differences with 95\% confidence
intervals, and partial $\eta^2$. With this sample the widest pairwise interval spans about 1.8 points, so adjusted differences smaller than about 0.9 points, half a residual standard deviation, could not be distinguished from zero; a difference of about 1.3 points, two thirds of a residual standard deviation, would be needed for 80\% power. Because we decided to
examine prior knowledge after inspecting the pre-to-post patterns, we
report that analysis as exploratory. The exploratory model allowed the effect of
pre-test score to differ by condition and kept pre-test score continuous
rather than split into groups, because splitting loses power and makes
results depend on where the cut
falls~\cite{maccallum2002dichotomization}. We report the
interaction test, the fitted lines, and condition differences at pre-test
scores of one and six, the sample's lower and upper quartiles. A median-split
version of this analysis appears in Appendix~\ref{app:median-split} as an exploratory check.

\subsubsection{Content Analysis of Tutoring Interactions}
\label{sec:process-method}
To characterize what learners asked for and what the tutor did, we
labeled every tutor message together with the learner message that
prompted it, in the context of its \emph{problem trajectory}: one
participant's conversation and recorded paper work on one problem,
including worked-solution review. We used directed content
analysis~\cite{hsieh2005approaches}, drawing on help-seeking and tutorial
dialogue
research~\cite{graesser1994question,chen2011exploring,shute2008formative}.
The codebook has two parts. \emph{Help-seeking purposes} (six labels)
describe what a learner's request sought. \emph{Tutor moves} (eight
labels) describe what the tutor's message did. A message can carry more than one label, because one message often does
several things at once. ``You dropped the negative sign'' gives correctness feedback and
identifies an error, and a single request can ask both for a check and
for the next step. Purpose labels apply only to learner messages that
request help, which the codebook identifies first; tutor moves apply to
every tutor message, requested or proactive.
Appendix~\ref{app:process-specification} defines the labels.

Three annotators from the research team developed and calibrated the
codebook. They independently annotated 39 messages from 16 trajectories,
discussed disagreements, refined the definitions, and re-annotated all 39
with the final codebook. They then independently annotated a held-out
evaluation set
of 155 messages from 32 further trajectories, sampled across
conditions. We measured agreement with Krippendorff's
$\alpha$~\cite{artstein2008agreement}. Because a message can carry
several purpose or move labels, we compared label sets using the MASI
distance~\cite{passonneau2006masi}, which credits partial overlap between
two annotators' sets rather than counting any difference as a full
disagreement, as in prior multi-label dialogue
annotation~\cite{ferschke2012behind,heddaya2023language}. Agreement among the annotators was high
($\alpha=.799$--$.917$; Table~\ref{tab:process-agreement}), indicating
that the codebook could be applied consistently.

Annotating all 2,140 tutor messages with their paper views by hand was
impractical, so we used GPT-5.6 Terra as an annotation model, following
recent work on human-evaluated LLM
annotation~\cite{xiao2023deductive,pangakis2025keeping,an2026coevaluation}.
The annotation model differed from the tutoring model, so the tutor did
not score its own messages.
The model received the same materials as the annotators: the codebook,
the conversation, the task and reference solution, and the recorded
paper views. We developed its prompts against the annotators'
annotations of the 39-message calibration set, froze them, and then
evaluated the model on the 155-message evaluation set against each
annotator (Table~\ref{tab:process-agreement}). Model--human agreement was high for purposes ($\alpha=.867$--$.928$) and lower for moves ($\alpha=.655$--$.693$). The move coefficient reflects three labels with weaker
agreement (task clarification, .47--.59; other, .59--.74; explanation,
.72--.74), on which we make no condition claims. For the five moves we
interpret in Section~\ref{sec:process-results}, per-label agreement with
each annotator was at least .76
(Table~\ref{tab:process-label-agreement}).
The model and the annotators agreed at $\alpha=.941$ on whether a learner
message requested help. The frozen model then
annotated the full corpus, which covers the 114 participants who received tutor messages. Appendix~\ref{app:calibration-development} gives the
sampling, annotation procedure, and per-label agreement.
The annotators also rated mathematical correctness in the same
evaluation set, and each rated 96.6--97.3\% of the 146 messages with a
mathematical claim as correct (Appendix~\ref{app:math-correctness}).

\begin{table*}[!htbp]
\centering
\caption{Agreement on the 155-message evaluation set. The \emph{Among annotators} column
reports agreement among the three human annotators; the remaining
columns compare one fixed model output with each annotator separately.
Both label sets use Krippendorff's $\alpha$ with a set-overlap distance
(MASI).}
\label{tab:process-agreement}
\small
\begin{tabular}{@{}lrrrr@{}}
\toprule
& Among & \multicolumn{3}{c}{Model versus} \\
\cmidrule(l){3-5}
Label set & annotators & Annotator 1 & Annotator 2 & Annotator 3 \\
\midrule
Help-seeking purposes & .917 & .903 & .928 & .867 \\
Tutor moves & .799 & .693 & .655 & .660 \\
\bottomrule
\end{tabular}
\end{table*}

We report each purpose as its share of a participant's help requests
and each move as its share of a participant's tutor messages, averaged
within condition with 95\% participant-bootstrap confidence intervals;
participants with no applicable message contribute no share. The reported
percentages are therefore means of participants' shares rather than
percentages of the pooled messages. We compared participant shares across conditions with Kruskal--Wallis
tests, Holm-corrected across the labels of each label set, with
Holm-corrected pairwise Mann--Whitney tests within a label. Each
participant contributes one share per label, so that talkative learners
do not dominate.
For \CAct{} we also report move shares separately for requested replies
and proactive messages (Appendix~\ref{app:process-tests}). These analyses
characterize assistance requested and provided, not learner uptake.

\subsubsection{Survey Ratings}
\label{sec:survey-method}
We examined self-reported perceptions of support for ongoing work,
control and proactive help, and camera setup through individual
final-survey items. ``Not applicable'' responses were excluded, and the
number of numeric responses is reported per item. We summarized each item
by condition as means and standard deviations. Holm correction was
applied within each item, not across items, and items shown only to the
two visual conditions were compared with a Mann--Whitney $U$ test.

\subsubsection{Thematic Analysis of Participant Accounts}
\label{sec:thematic-method}
The final survey ended with two open questions. The first asked what
participants had wanted from the tutor during the activity, including
what worked and what they would change; the second asked how they would
feel about tackling a similar problem without it. An optional third field invited further
feedback, which 59 participants used.
Responses to the two questions were predominantly spoken. Across the two
questions, the material comprises 180 transcribed recordings averaging 139 words and 52 typed answers, plus 59 optional comments: about 28,500 words from all 117
participants.\footnote{Two additional recordings from one participant
could not be transcribed; that participant's optional typed comment was
included.}
One researcher conducted a reflexive thematic
analysis~\cite{braun2006thematic,braun2024rtarg,braun2021survey}
of these accounts, coding close to participants' own wording. Each participant's responses were read together to
preserve the context of their account. The researcher coded the accounts and grouped related codes into
candidate themes. Each theme was then checked against its coded passages and complete accounts, and its central idea and boundaries were
refined using contrasting and ambivalent cases.
Condition comparisons were made only after candidate themes were
established.

\section{Results}
\label{sec:results}

We report results for each research question in turn.

\subsection{Tutor Use, Problems Attempted, and Learning (RQ1)}
\label{sec:quantitative-results}

We did not detect a difference between conditions in tutor use or in
problems attempted. Learners gained from pre-test to post-test in every
condition, and no significant difference emerged in adjusted post-test
scores.

\subsubsection{Tutor Use by Condition, Mostly Learner-Initiated}
\label{sec:tutor-use-results}
Most dialogue occurred during problem solving, with little exchange
during worked-solution review (Table~\ref{tab:process-help-frequency}).
Tests of message volume were not significant for learner messages or
for tutor messages, and the medians were close. In \CAct{}, the tutor sent 4.2 proactive messages per participant in addition to 16.2 requested replies, versus 18.8 requested replies in \CBase{} and 15.5 in \CSee{}. Tutoring was predominantly learner-initiated: even in \CAct{}, about 80\% of tutor messages were requested replies.

\begin{table*}[!htbp]
\centering
\caption{Learner messages, tutor messages, and problems attempted by condition. Entries are participant means (SD); medians in brackets and Kruskal--Wallis tests are given for the four tested rows. The problem-solving and worked-solution-review rows are descriptive breakdowns. Em dashes mark a measure not applicable to the reactive conditions. $\epsilon^2$ is not bias-corrected; its expected value with no effect at this sample size is about .02.}
\label{tab:process-help-frequency}
\setlength{\tabcolsep}{2.5pt}\small
\begin{tabularx}{\linewidth}{Xrrrl}
\toprule
Measure & \CBase{} & \CSee{} & \CAct{} & Test \\
$n$ & 40 & 37 & 40 & \\
\midrule
\multicolumn{5}{l}{\textit{Across the assisted activity}} \\
Learner messages & 19.08 (15.46) [16.5] & 15.68 (10.50) [14] & 16.53 (16.41) [13] & $H(2)=1.4$, $p=.499$, $\epsilon^2=.012$ \\
Tutor messages (total) & 18.78 (15.27) [16.5] & 15.49 (10.52) [14] & 20.40 (15.19) [16] & $H(2)=1.3$, $p=.516$, $\epsilon^2=.011$ \\
\quad Requested replies & 18.78 (15.27) [16.5] & 15.49 (10.52) [14] & 16.23 (16.25) [12.5] & $H(2)=1.4$, $p=.501$, $\epsilon^2=.012$ \\
\quad Proactive messages & --- & --- & 4.18 (3.64) & \\
\addlinespace[3pt]
\multicolumn{5}{l}{\textit{During problem solving}} \\
Learner messages & 17.78 (15.09) & 14.35 (10.79) & 15.60 (16.13) & \\
Tutor messages & 17.53 (14.87) & 14.27 (10.77) & 19.50 (15.05) & \\
\addlinespace[3pt]
\multicolumn{5}{l}{\textit{During worked-solution review}} \\
Learner messages & 1.30 (2.88) & 1.32 (1.99) & 0.93 (1.51) & \\
Tutor messages & 1.25 (2.86) & 1.22 (1.81) & 0.90 (1.35) & \\
\addlinespace[3pt]
Problems attempted (of 10) & 5.65 (2.14) [5] & 6.05 (2.31) [5] & 6.23 (2.15) [5] & $H(2)=1.5$, $p=.475$, $\epsilon^2=.013$ \\
\bottomrule
\end{tabularx}
\end{table*}

\subsubsection{Problems Attempted by Condition}
\label{sec:completion-results}
Participants attempted about six of the ten problems on average
(Table~\ref{tab:process-help-frequency}). The comparison between
conditions was not significant. In the exploratory model described in Section~\ref{sec:quantitative-method}
(interaction $F(2,111)=1.00$, $p=.370$, partial $\eta^2=.018$), moderation
by prior knowledge was likewise not detected.

\subsubsection{Every Condition Improved; Adjusted Scores by Condition}
\label{sec:learning-results}
Across the whole sample, scores rose from 3.3 to 4.6 of nine, a
significant increase of moderate size, and every condition improved
(Table~\ref{tab:learning}). Adjusting for pre-test score, we did not
detect a difference between conditions. The largest adjusted difference,
two thirds of a point favoring \CAct{} over \CSee{}, was about a third of
a residual standard deviation and a quarter of the post-test standard
deviation, with an interval that includes zero.

\begin{table*}[!htbp]
\centering\small
\setlength{\tabcolsep}{3pt}
\caption{Learning by condition. Entries are means (SD); brackets are 95\% confidence intervals. Gain is post-test minus pre-test. Adjusted post-test scores and the pairwise differences below come from an ANCOVA with continuous pre-test score as covariate, evaluated at the sample mean. The condition test, the intervals, and the contrasts use robust (HC3) standard errors; partial $\eta^2$ is from the ordinary least-squares fit. The third column of the lower table is the adjusted difference divided by the residual SD (1.86); $p$-values are Holm-corrected across the three contrasts.}
\label{tab:learning}
\begin{tabularx}{\linewidth}{Xrrrll}
\toprule
Condition & $n$ & Pre-test (of 9) & Post-test (of 9) & Gain [95\% CI] & Adjusted post-test [95\% CI] \\
\midrule
\CBase{} & 40 & 2.98 (2.67) & 4.40 (2.85) & 1.43 [0.90, 1.95] & 4.68 [4.15, 5.20] \\
\CSee{} & 37 & 3.70 (3.00) & 4.49 (2.97) & 0.78 [0.12, 1.45] & 4.22 [3.59, 4.84] \\
\CAct{} & 40 & 3.38 (2.54) & 4.90 (2.48) & 1.52 [0.81, 2.24] & 4.88 [4.23, 5.52] \\
\midrule
\multicolumn{6}{@{}l}{\footnotesize Whole sample: 3.34 to 4.60, paired $t(116)=6.85$, $p<.001$, $d=0.63$. Condition (robust): $F(2,113)=1.18$, $p=.310$; partial $\eta^2=.022$.} \\
\bottomrule
\end{tabularx}
\par\smallskip
\begin{tabularx}{\linewidth}{Xlrr}
\toprule
Adjusted pairwise difference & Difference [95\% CI] & Diff.\ / resid.\ SD & Holm $p$ \\
\midrule
\CAct{} vs.\ \CBase{} & 0.20 [$-$0.64, 1.04] & 0.11 & $.640$ \\
\CAct{} vs.\ \CSee{} & 0.66 [$-$0.23, 1.55] & 0.35 & $.429$ \\
\CSee{} vs.\ \CBase{} & $-$0.46 [$-$1.27, 0.35] & $-$0.25 & $.526$ \\
\bottomrule
\end{tabularx}
\end{table*}

\begin{figure}[!t]
\centering
\includegraphics[width=\linewidth,alt={Scatter plot of post-test versus pre-test score, both out of nine, with one fitted regression line and confidence band per condition. The Text-Reactive line is steepest and the Visual-Proactive line flattest, so Visual-Proactive is highest at low pre-test scores and lowest at high scores; the lines cross between pre-test scores of about four and seven. A dotted diagonal marks no change.}]{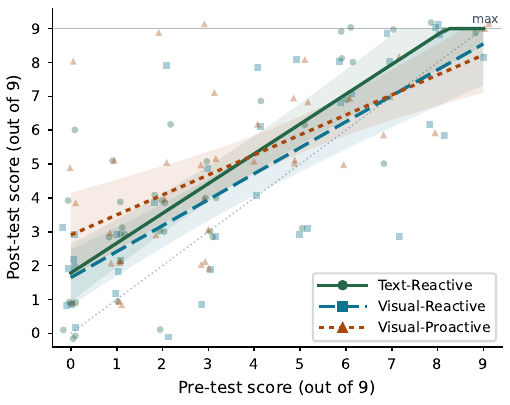}
\caption{Exploratory moderation by prior knowledge: post-test versus
pre-test score with fitted lines and 95\% confidence bands per condition.
Points are jittered; the gray dotted diagonal marks no change, and fitted
values are clipped at the maximum score of nine. Interaction (OLS) $F(2,111)=1.63$, $p=.201$, partial $\eta^2=.028$.}
\Description{A scatter plot with pre-test score on the horizontal axis
and post-test score on the vertical axis, each from zero to nine. Points
for the three conditions are shown with distinct markers and colors.
Three fitted lines with shaded confidence bands rise from lower left to
upper right. The Visual-Proactive line starts highest near a pre-test of
zero and rises least steeply; the Text-Reactive line starts lower and
rises most steeply; the Visual-Reactive line starts lowest and ends
between the other two. The Visual-Proactive line crosses Text-Reactive
near a pre-test score of four and Visual-Reactive near seven. A dotted diagonal shows where post-test equals pre-test.}
\label{fig:learning-moderation}
\end{figure}

\paragraph{Exploratory: no significant moderation by prior knowledge.}
The pre-test $\times$ condition interaction was not significant
(Figure~\ref{fig:learning-moderation}), though the fitted lines were not
parallel. At a pre-test score of one, \CAct{} exceeded \CSee{} by 1.08
points [$-$0.18, 2.35] and \CBase{} by 0.83 [$-$0.42, 2.09]; at six the
differences were 0.20 [$-$0.81, 1.21] and $-$0.61 [$-$1.56, 0.34]. The lines crossed between pre-test scores of about four and
seven. A median-split version of this
analysis (Appendix~\ref{app:median-split}) yielded one nominally
significant contrast that did not hold up under robust standard errors,
correction across contrasts, or a one-point change in the cut point. We
therefore report a possible advantage of proactive tutoring for learners
with low prior knowledge as a hypothesis for a larger study, not a
finding.

\subsection{Help Seeking and Tutor Assistance (RQ2)}
\label{sec:process-results}

RQ2 asks what the assistance consisted of: what learners sought from
the tutor (\emph{help-seeking purposes}) and what the tutor did
(\emph{tutor moves}).
Figure~\ref{fig:process-labels} compares them across conditions, using
labels from the validated annotation model
(Section~\ref{sec:process-method}): purposes cover every help request,
and moves cover every tutor message, requested or proactive. Six of the 40 \CAct{} participants and three of the 37 in \CSee{} never asked for help,
versus none in \CBase{}, so they contribute no purpose share. Five of the six in \CAct{} received proactive messages without ever writing to the tutor. Percentages are means of participant shares (Section~\ref{sec:process-method}). Because a message can carry more than
one label, percentages need not sum to 100.
Tests are Kruskal--Wallis comparisons of participant shares,
Holm-corrected within each label set (Table~\ref{tab:process-tests} in
Appendix~\ref{app:process-tests}); asterisks in
Figure~\ref{fig:process-labels} mark labels that differ significantly
after correction.

\subsubsection{With a Paper View, the Tutor Confirmed Work Instead of Asking for It}
\label{sec:visual-results}

\begin{figure*}[!t]
\centering
\includegraphics[width=.72\linewidth,alt={Two dot plots. Panel a shows six help-seeking purposes and panel b shows eight tutor moves. Each row gives the mean percentage of participants' help requests or tutor messages carrying that label for Text-Reactive, Visual-Reactive, and Visual-Proactive, with 95 percent confidence bands and the means printed at the right. In panel a, verification is higher and procedural guidance lower in the two visual conditions, but no purpose differs significantly. In panel b, correctness feedback is higher in the visual conditions, error identification and directive hints are highest in Visual-Proactive, and elicitation is lower in the visual conditions; these four moves are marked with asterisks.}]{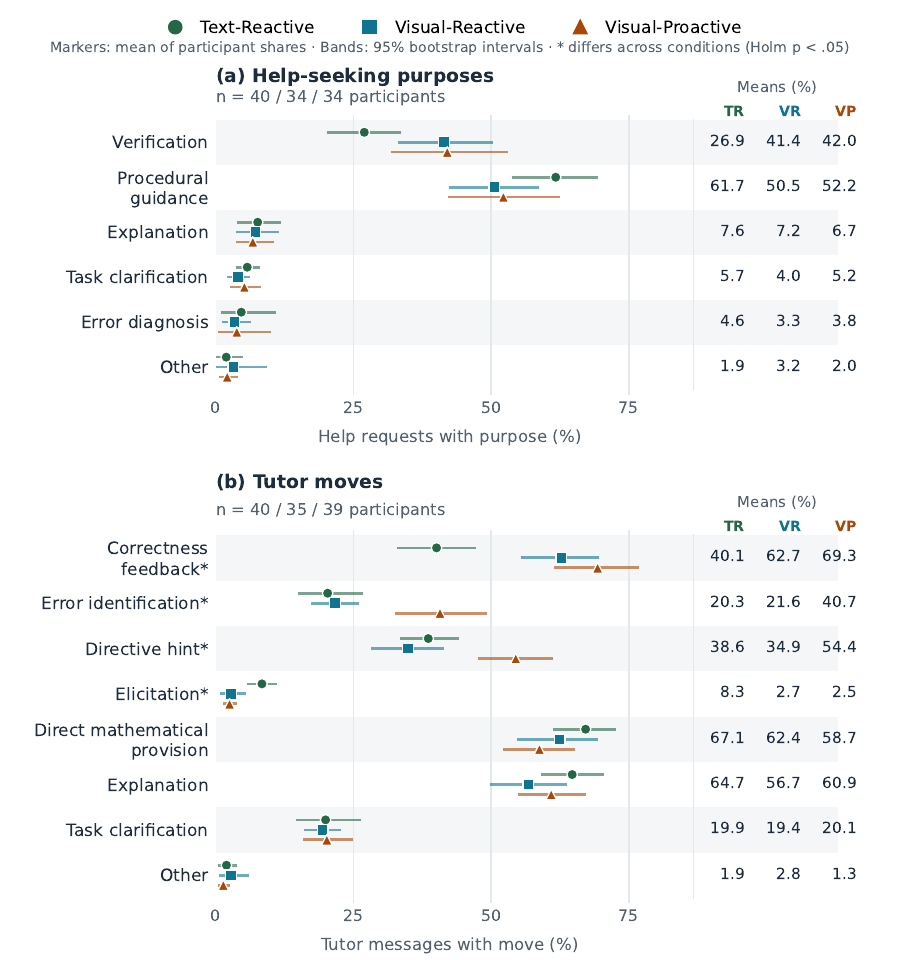}
\caption{Help-seeking purposes (a) and tutor moves (b) by condition.
Markers are means of participant shares; panel $n$ and the means are printed in the figure. The corpus is 602/505/562 help requests and 751/573/816 tutor
messages in \CBase{} (TR), \CSee{} (VR), and \CAct{} (VP), including
167 proactive messages. Labels are ordered as discussed in the text;
per-label tests are in Table~\ref{tab:process-tests}.}
\Description{Panel a has six rows comparing the three conditions on
help-seeking purposes. Verification has larger mean shares in the visual
conditions and procedural guidance a larger share in Text-Reactive; the
remaining purposes have overlapping intervals, and no row carries an asterisk. Panel b
has eight rows on tutor moves. Correctness feedback has larger mean
shares in the visual conditions; error identification and directive
hints are highest in Visual-Proactive; elicitation is lower in the
visual conditions; these four rows carry asterisks. Explanation, direct
provision, task clarification, and other have overlapping intervals and
carry no asterisk.}
\label{fig:process-labels}
\end{figure*}

\begin{figure*}[!t]
\centering
\begin{minipage}[t]{.48\linewidth}
\small\raggedright
\textbf{(a) \CSee{}: verification}\par\vspace{4pt}
\parbox[c][120pt][c]{\linewidth}{\centering
\includegraphics[trim=180bp 570bp 100bp 530bp,clip,angle=180,origin=c,
width=\linewidth,totalheight=118pt,keepaspectratio,alt={Handwritten right triangle and constraint L squared equals x squared plus 3 squared, with a rope-shortening rate of minus one meter per second. The accompanying learner message asks whether this work is right.}]{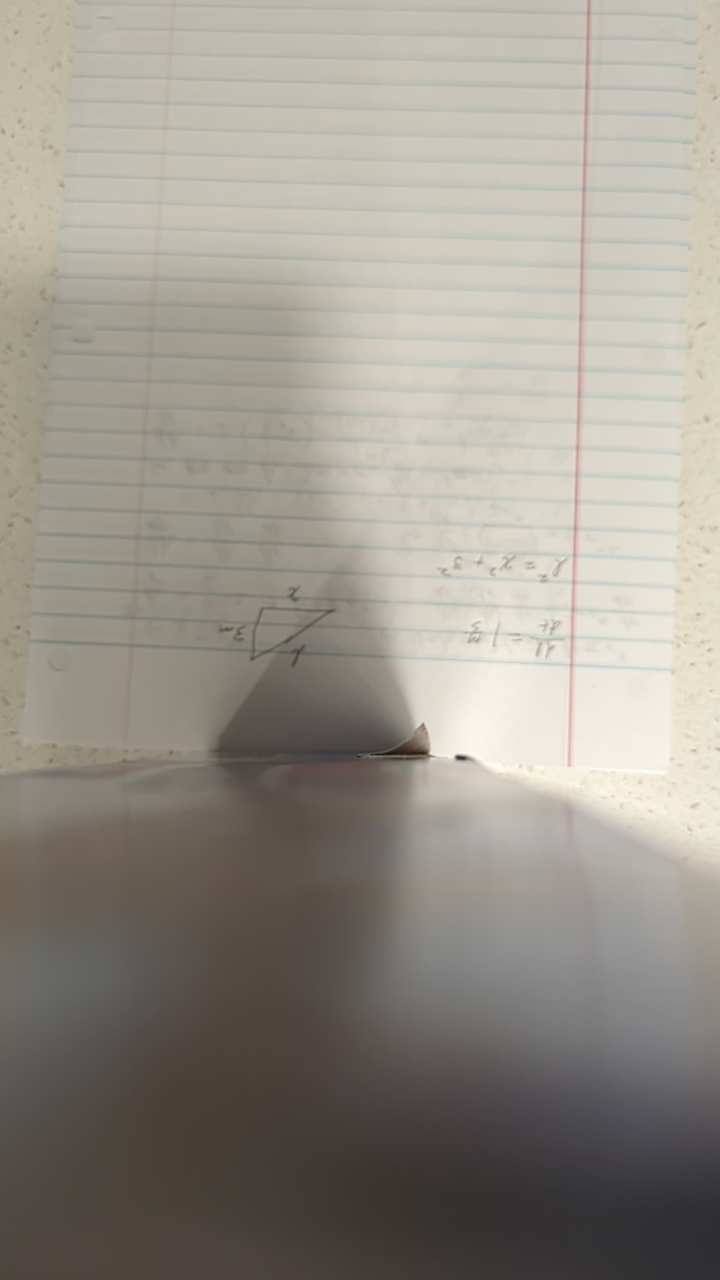}}
\par\vspace{5pt}
\textbf{Learner:} ``am I right so far?''\par\vspace{4pt}
\textbf{Tutor:} ``Yes. Your triangle and constraint
$L^2=x^2+3^2$ are correct, and $dL/dt=-1\text{ m/s}$ correctly
represents the rope shortening.''
\end{minipage}\hfill
\begin{minipage}[t]{.48\linewidth}
\small\raggedright
\textbf{(b) \CAct{}: proactive repair}\par\vspace{4pt}
\parbox[c][120pt][c]{\linewidth}{\centering
\includegraphics[trim=290bp 460bp 137bp 365bp,clip,angle=90,origin=c,
width=\linewidth,totalheight=118pt,keepaspectratio,alt={Handwritten differentiation of length times width, omitting width from the term containing the rate of change of length. The accompanying tutor message identifies the missing factor and prompts use of the product rule.}]{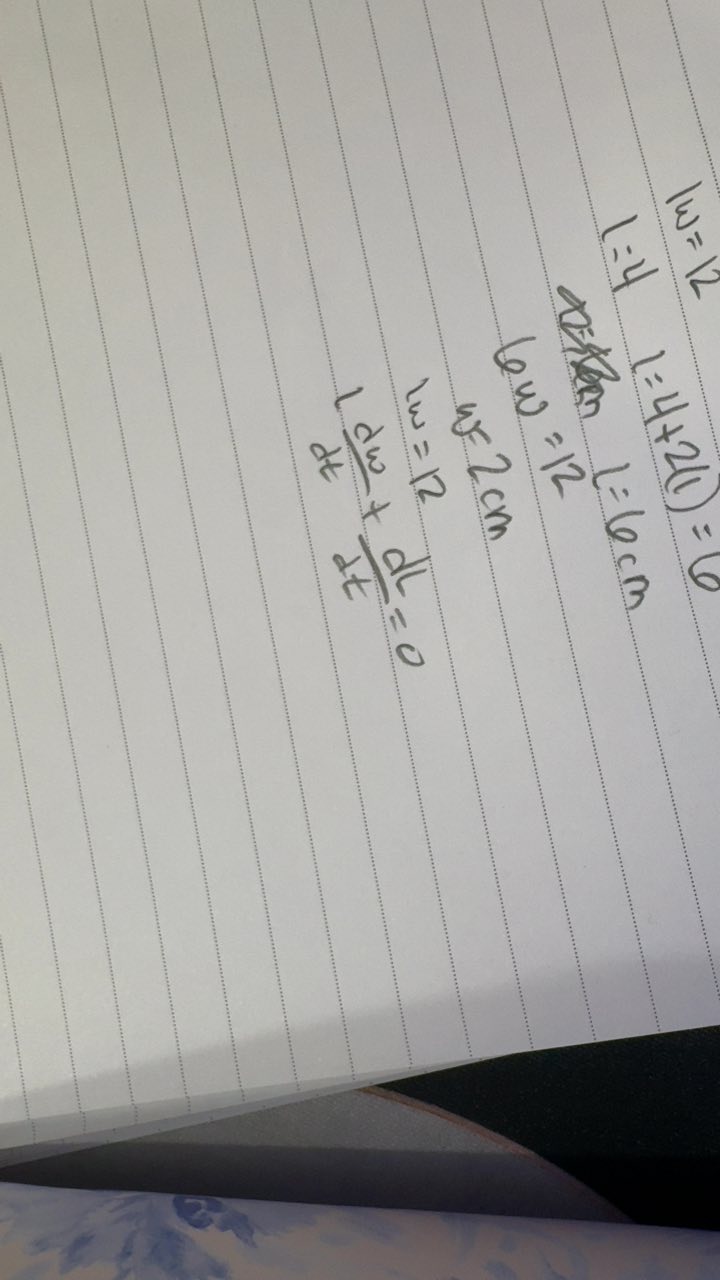}}
\par\vspace{5pt}
\textbf{Tutor-initiated; no learner request.}\par\vspace{4pt}
\textbf{Tutor:} ``In your derivative of $\ell w$, the term containing
$d\ell/dt$ is missing a factor. Since both dimensions vary, multiply each
dimension's rate by the other undifferentiated dimension.''
\end{minipage}
\caption{Assistance referring to handwritten work. (a) P31 asked for
verification without restating the triangle, constraint, or rate on the
page. (b) A proactive message to P04 identified a missing factor in a
differentiated constraint and hinted at the product rule. Paper views are
those supplied to the tutor, cropped and rotated; excerpts are verbatim.}
\Description{Two paper crops alongside their corresponding dialogue.
In the first, the learner has drawn a right triangle, written
L squared equals x squared plus 3 squared, and recorded a rope rate of
minus 1 meter per second. The tutor confirms this work in response to
a short verification request. In the second, the derivative of lw
is written as l times dw/dt plus dl/dt equals zero, omitting w from
the second term. The proactive tutor identifies the missing factor
and reminds the learner of the product rule without supplying the
corrected equation.}
\label{fig:process-examples}
\end{figure*}

Learners' requests leaned toward asking the tutor to check work, and
the tutor's messages changed more clearly still.
Verification requests ask the tutor to check an answer, expression, or
approach. They made up just over 40\% of requests in both visual
conditions and 27\% in \CBase{} (Figure~\ref{fig:process-labels}a), a
medium-sized difference that did not survive correction across the six
purposes. Procedural-guidance requests, which ask for a
formula, method, or next step, were lower in the visual conditions, at
about half of requests versus 62\%, though not significantly. The
tutor's messages showed the matching pattern, and here the differences
were large and significant. Correctness feedback affirms or rejects
existing work. Its share rose from 40\% of tutor messages in \CBase{} to
63\% in \CSee{} and 69\% in \CAct{} (59\% among \CAct{}'s requested replies, Table~\ref{tab:process-decomposition}; Figure~\ref{fig:process-labels}b). Elicitation, asking learners
to show their work, fell from 8\% of tutor messages to under 3\%.
On the boat-and-rope problem, a \CBase{} participant asked ``how to
approach'' and received the relationship $L^2=x^2+3^2$. A \CSee{}
participant (P31; Figure~\ref{fig:process-examples}a) opened with ``am I
right so far?'' and the tutor confirmed the triangle, the constraint, and the negative rope rate already on the
page, none of which P31 had typed. The paper view supplied what the tutor
would otherwise have had to request.

\subsubsection{Proactivity Added Repair Messages Without a Detected Change in Requests or Requested Replies}
\label{sec:proactive-results}

Adding initiative changed the assistance learners received. We did not detect a difference in purpose shares between \CAct{} and \CSee{}; the intervals overlapped on every label (Figure~\ref{fig:process-labels}a). In \CAct{}, error identification, naming a specific mistake, appeared in 41\% of a participant's tutor messages on average, versus 20\% in \CBase{} and 22\% in \CSee{}. Directive hints, pointing to the next action, appeared in 54\%, versus 39\% and 35\%
(Figure~\ref{fig:process-labels}b). Both differences were significant
(Table~\ref{tab:process-tests}). We did not detect a difference in direct
mathematical provision, among the most common moves in every condition,
and we do not interpret the three labels with weaker agreement. The two
significant differences came from the proactive messages: when \CAct{} is
represented by its requested replies alone, the differences were not detected (Table~\ref{tab:process-decomposition} in Appendix~\ref{app:process-tests}). Explicit
error-diagnosis requests were rare in every condition, so the diagnosis
was the tutor's own contribution. Figure~\ref{fig:process-examples}b
shows one such proactive message to P04. It identified a missing factor
in the handwritten derivative of $\ell w$ and reminded the learner to
multiply each dimension's rate by the other, leaving the revision to them.

\subsubsection{The Proactive Tutor Spoke Rarely and Mostly to Repair Errors}
\label{sec:proactive-decisions}

Across the 40 \CAct{} participants, the tutor completed 5,047 paper analyses, a median of 127 per participant, and after 96\% of them it chose to wait. It decided to intervene 187 times and delivered 167 proactive messages to 32 participants, withholding 20 because the learner sent a chat message while the analysis was running; eight participants received no proactive message. Nearly every delivered
message was a repair of a visible error (157). Transition support for a
stalled approach, the second intervention type in
Section~\ref{sec:tutoring-policies}, was delivered only 10 times. These decisions explain the profile in
Figure~\ref{fig:process-labels}b: because nearly every proactive message
was a repair, proactive messages carried correctness feedback, error identification, and directive hints far more often than requested replies did (Table~\ref{tab:process-decomposition}).

\subsection{Learner Experience (RQ3)}
\label{sec:experience-results}

Learners rated the visual conditions more highly on support for their
ongoing work, and their own accounts explain what they valued and what
they still wanted. We report the survey ratings first, then the four
themes from the open responses.

\subsubsection{Ratings Favored Visual Access; Perceived Control Was High Throughout}
Learners in both visual conditions agreed more strongly than \CBase{}
learners that they received useful help without restating their written
steps (Table~\ref{tab:survey-ratings}). \CSee{} learners also rated the fit of feedback to their current solving
stage above \CBase{}, with \CAct{} not distinguishable from either. Learners in both visual conditions agreed that feedback accurately
reflected their written work, with \CSee{} rated significantly higher ($p=.045$; Table~\ref{tab:survey-ratings}). Agreement that learners controlled which step to try next
was high in every condition, and the test across conditions was not significant. The \CAct{} learners who received proactive help agreed that it
supported their progress and disagreed that it interrupted their work. Camera
setup was rated moderately easy in all conditions. Ratings tracked the
content analysis:
visual access improved perceived fit to ongoing work.

\begin{table*}[!htbp]
\centering\footnotesize
\setlength{\tabcolsep}{3pt}
\renewcommand{\arraystretch}{1.2}
\caption{Survey ratings of tutoring and camera setup by condition. Entries are means (SD) with medians in brackets on a scale from 1 (strongly disagree) to 7 (strongly agree); $n$ counts numeric responses per condition in column order. The two proactive-help items are summarized for the \CAct{} participants who received at least one proactive message. Em dashes mark items not administered or not tested. Within a row, means that do not share a superscript letter differ in Holm-corrected pairwise Mann--Whitney tests; the two-condition item has a single test.}
\label{tab:survey-ratings}
\begin{tabularx}{\linewidth}{>{\raggedright\arraybackslash}p{.30\linewidth}cccc>{\raggedright\arraybackslash}X}
\toprule
Item & \CBase{} & \CSee{} & \CAct{} & $n$ & Test \\
\midrule
\multicolumn{6}{l}{\textit{Support for ongoing work}} \\
Useful help without restating written steps & 4.21\textsuperscript{a} (1.85) [5] & 5.57\textsuperscript{b} (1.60) [6] & 5.15\textsuperscript{b} (1.61) [5] & 39/35/40 & $H(2)=12.2$, $p=.002$, $\epsilon^2=.108$ \\
Feedback matched current solving stage & 5.18\textsuperscript{a} (1.53) [5.5] & 6.06\textsuperscript{b} (1.45) [7] & 5.88\textsuperscript{ab} (1.09) [6] & 40/35/40 & $H(2)=10.7$, $p=.005$, $\epsilon^2=.094$ \\
Feedback accurately reflected written work & --- & 5.92\textsuperscript{a} (1.52) [6.5] & 5.50\textsuperscript{b} (1.24) [6] & ---/36/40 & $U=905.5$, $p=.045$, $r=.26$ \\
\addlinespace[3pt]
\multicolumn{6}{l}{\textit{Control and proactive guidance}} \\
Felt in control of which step to try next & 5.10 (1.82) [6] & 5.86 (1.00) [6] & 5.65 (1.12) [6] & 39/37/40 & $H(2)=2.6$, $p=.277$, $\epsilon^2=.022$ \\
Proactive help supported progress & --- & --- & 5.53 (1.32) [6] & 32 & --- \\
Proactive messages interrupted work & --- & --- & 2.45 (1.46) [2] & 31 & --- \\
\addlinespace[3pt]
\multicolumn{6}{l}{\textit{Camera setup}} \\
Camera setup was easy & 5.10 (1.58) [6] & 4.59 (1.64) [5] & 5.13 (1.67) [5] & 40/37/40 & $H(2)=2.6$, $p=.268$, $\epsilon^2=.023$ \\
\midrule
\multicolumn{6}{@{}p{\linewidth}@{}}{\footnotesize Holm-corrected pairwise tests behind the letters, with rank-biserial $r$: useful help, \CSee{} vs.\ \CBase{} $p=.003$, $r=.44$; \CAct{} vs.\ \CBase{} $p=.040$, $r=.30$; \CSee{} vs.\ \CAct{} $p=.167$. Feedback matched stage, \CSee{} vs.\ \CBase{} $p=.004$, $r=.41$; \CAct{} vs.\ \CBase{} $p=.093$; \CSee{} vs.\ \CAct{} $p=.175$.} \\
\bottomrule
\end{tabularx}
\end{table*}

\subsubsection{Participant Accounts}
\label{sec:thematic-results}
Four themes summarize the 117 open-response accounts
(Section~\ref{sec:thematic-method}). Counts are indicative of prevalence
in one researcher's coding: they give the number of participants whose
accounts expressed an idea, out of 117 unless a smaller group is named.
Accounts commonly expressed several ideas, so counts overlap. Spoken
responses were machine-transcribed; we normalized capitalization in
quotations and marked omissions with ellipses. Quotations otherwise
preserve the original wording.

\paragraph{Verification in place, with a risk of reliance.}
Learners in the visual conditions valued a tutor that read their work
and checked it in place, echoing the rise in verification requests
in Section~\ref{sec:visual-results}. Of the 77 participants with a paper
view, 12 said the tutor understood what their writing referred to and 6
described being spared from retyping. P89 (\CSee{}) valued confirmation that came ``just by
looking at my work so I didn't have to like type it in,'' and
P30 (\CAct{}) found that ``being able to show it where I'm at instead of
having to try to write out that problem into a text based answer is so
much easier.'' Checking reassured 21 of the 77, compared with 3 of the 40 in \CBase{},
and it mattered even to learners who felt able to solve alone: P14 (\CSee{}) could now solve similar problems ``but it
would be considerably more difficult without knowing if I have gone
wrong on a particular step.'' Several \CBase{} learners wanted the same;
P02 (\CBase{}) ``would have liked for the AI to be able to see what I'm doing on
the paper and read the steps that I'm doing and follow along.'' Easy
checking had a cost for some: 6 participants, all in the visual
conditions, worried about leaning on the tutor, and P57 (\CSee{})
admitted to ``relying on it a bit too much at times I could have probably tried and
remembered some of these things on my own.''

\paragraph{Demand for more initiative.}
Asking for help was itself work for 23 participants, and 20 wished the
tutor would step in without being asked. P55 (\CSee{}) ``wouldn't even know how I
should phrase my question to have the AI give a useful response,'' and
P77 (\CBase{}) said that ``if I would be stuck somewhere I rather would
like it to push me without me asking.'' Most of the 20 were in \CAct{}
(12, versus 8 in the two reactive conditions), and 11 of those 12 (of the 32 recipients) had
already received proactive help, a median of 6 messages each, and still
wanted more. P54 (\CAct{}), who received
10 proactive messages, ``would have preferred if I was having a lot of trouble that they give me hints,
instead of waiting for me to ask for help.'' Recipients also valued the corrections themselves: P26 (\CAct{}) liked that ``it worked
without me having to ask a question because I didn't realize the first
one I did something wrong.'' The demand had a boundary that matched the
repair-first policy. For P17 (\CAct{}), ``more of the unsolicited
\ldots{} assistance would be nice but not too much more because you
still want to feel like you're working through it on your own,'' and
P108 (\CAct{}) drew the line by purpose: the tutor should ``only point
out mistakes unprompted rather than pointing out the next step
unprompted.'' By their own accounts, learners wanted the tutor to keep intervening
on errors, and more often.

\paragraph{Demand for explanation, with the step left to the learner.}
For 21 participants, the tutor's reminders brought back methods they
already knew. P26 (\CAct{}) found that ``reminding me of like the chain
rule product rule and how to apply those \ldots{} that was really
helpful,'' and P65 (\CBase{}) ``went from not even remembering the power
rule existed to now understanding.'' The most common criticism was that
explanations stopped short: 27 wanted fuller explanation and 21 wanted
more concrete steps. P11 (\CSee{}) ``wished the AI explained more on why
certain steps are done,'' and P82 (\CBase{}) found ``a lot of text but
not a lot of values.'' At the same time, 16 wanted to keep doing the
steps themselves. P49 (\CAct{}) would have ``the AI prompt me to try
each step myself first before offering the full mathematical
breakdown,'' and P88 (\CSee{}) appreciated that the tutor ``didn't baby
me or put the training wheels on too hard.'' These requests pull in different directions; P49 proposed prompting each step first and explaining afterward.

\paragraph{Mixed readiness after one session.}
Asked how they would feel tackling a similar problem without the tutor, 14
participants reported feeling more capable. P42 (\CBase{}) felt ``more
prepared because of what I learned from the AI tutor.''
Another 18 made their readiness conditional on the problem's difficulty
or on having enough time, and 12 said they needed more practice first; P93
(\CSee{}) still thought ``I need more practice before I could do this
without help.'' A larger group, 28 participants spread across the three
conditions, still described feeling lost or unable to start on their
own, and 8 reported no benefit from the tutor; P11 (\CSee{}) ``would feel
very lost as I wouldn't know where to start.'' A further 12 said they could manage alone but would rather keep the
tutor as a check; P63 (\CBase{}) ``probably could tackle the questions
alone, with much more time and error. But I would not want to if given
the choice.'' These accounts describe perceived readiness after one session; measured scores rose significantly in every condition (Section~\ref{sec:learning-results}).
Across the four themes, learners wanted a tutor that saw and checked their work, pointed out
errors before being asked, and explained why a step works.

\FloatBarrier
\section{Discussion}
\label{sec:discussion}

In a randomized study with 117 adults, a paper view changed what the tutor said, and permission to act changed what help arrived.

\subsection{Seeing the Work Changes What Has to Be Said}

A paper view removed a step from the interaction. The
tutor confirmed and corrected work the learner had never typed, and its
requests that learners communicate their work or reasoning fell to under 3\% of messages.
Learners' own requests leaned toward asking for a check, although the
corrected test was not significant
(Section~\ref{sec:visual-results}). Participants' accounts suggest why:
the page carried context that would otherwise be typed, and confirmation
of a correct step let them build on it.  The page appears to have served as common ground~\cite{clark1991grounding}
and performed both functions of a shared visual space, anchoring what
was said and showing the state of the task~\cite{gergle2013grounding}.
 A paper view brings to conversational
tutoring the feedback on intermediate work that step-based systems
provided~\cite{vanlehn2006behavior,vanlehn2011relative}, without their
structured input and without the single snapshot that photo-based apps
read~\cite{sloanlynch2022photoapps}. One design implication is that a
tutor with a view could offer to check the current step before being
asked.

\subsection{From Seeing to Acting: What Proactivity Added}

The added error identification and directive hints came from the proactive messages and were not detected when \CAct{} was represented by its requested replies alone (Section~\ref{sec:proactive-results}). Error diagnosis was among the
rarest request purposes, and P26 valued help that arrived before they had noticed their own mistake. Human tutors watch for such
errors~\cite{merrill1992effective,chi2001learning}; learning during
tutoring concentrates at impasses~\cite{vanlehn2003impasse}, and
learners often do not ask for help when they need
it~\cite{aleven2003helpseeking,roll2011helpseeking}.
Mixed-initiative design weighs the value of acting against the cost of
interrupting~\cite{horvitz1999mixedinitiative}, and the repair-first
policy applied that rule conservatively. It waited after 96\% of its
analyses and sent nothing at all to eight of the 40 participants. Participants who received proactive messages rated them supportive rather than interrupting, and 11 of the 12 who
wished for more had already received some. Prior proactive assistants
were preferred only when interjections were spaced out and limited~\cite{chen2025needhelp}, were sometimes experienced as interruptions when triggered by inactivity~\cite{pu2025codellaborator}, or left learners
feeling less in control~\cite{otmani2026proactive}. The messages here were few, about four per learner, which alone may explain their
acceptance. The trigger may also matter: the tutor acted on evidence of an error rather than on a pause, which may be part of why its messages were welcome.

\subsection{Three Decisions Behind Proactive Help}

Learners did not express one stable preference for more or less help.
The same learner could welcome a proactive error alert while asking
that next steps be left for them to request, and the accounts pulled in
different directions (Section~\ref{sec:thematic-results}). Contingent
scaffolding has long distinguished when to help, what to help with, and
how much of the task to take
over~\cite{wood1976tutoring,vandepol2010scaffolding}. Asking is work of
its own, and generative AI makes it more demanding still (Section~\ref{sec:related-proactive})~\cite{tankelevitch2024metacognitive}; 23 participants described
asking for help in these terms. A single ``more or less proactive''
setting conflates when to help, what to help with, and how much to take over, and the accounts distinguish them.
Learners wanted the tutor to intervene when they were wrong or stuck;
they divided on how much of a step it should perform (16 wanted to keep
the steps, 21 wanted more concrete steps) and on how deeply it should
explain. The waiting rules and the model's error threshold kept help from crowding the conversation, and some learners would have welcomed it more often. Tutor policies could therefore set the timing, focus, and depth of
proactive help separately, with each choice evaluated against learning
alongside learner preference~\cite{koedinger2007assistance}. Because the three dimensions
were fixed together in a single policy, their separation here rests on
learners' accounts rather than on an experimental contrast.

\subsection{Learning After One Session}

Learners improved significantly from pre-test to post-test in every condition. Adjusted post-test scores did not differ detectably between conditions, and an exploratory trend favored the proactive tutor for learners with low pre-test scores (Section~\ref{sec:learning-results}). The
study was not powered to detect an interaction of the observed size
(Section~\ref{sec:limitations}), so the analysis does not rule out the
pattern the assistance dilemma~\cite{koedinger2007assistance} predicts,
in which more help benefits learners with low prior knowledge more than
those with high prior knowledge. The added proactivity was also light: about
four proactive messages per learner, and learning was assessed with a nine-item test.

\subsection{Everyday Devices as a Tutoring Interface}
\label{sec:deployment-discussion}

Ordinary paper and a phone lower the hardware requirements of visual
tutoring relative to tablets and digital
canvases~\cite{liu2026amiwrite,lee2025sketchpad}. They let a
population that mostly preferred paper keep writing where they already
write. The remote study also exposed the cost of the setup: 74 people declined after seeing the requirements, and about three in ten of those who entered exited during camera setup, most because the camera would not connect or the view would not stay readable (Section~\ref{sec:participants}). Connection failures suggest software or device problems; stability problems are mechanical, and an inexpensive stand
might reduce that source of exits in a deployment. Kraut et
al.~\cite{kraut2003visual} report a comparable cost in remote
collaboration, where aiming the camera took effort of its own.
Participants who completed the study rated setup moderately easy, and those ratings come only from people who established a view. Cloud
inference carries a per-session cost, higher in the proactive
condition, and sends written work to a third party. Smaller models
specialized for handwritten mathematics, run locally, could reduce both
cost and privacy exposure, as has been demonstrated for programming
support~\cite{kotalwar2024hints}.

\subsection{A Dataset for Studying Assistance Around Ongoing Work}
\label{sec:dataset}

Because the platform recorded the paper in all three conditions, the
dataset supports comparisons that one tutor's logs cannot. Trajectories
can be replayed to a different tutor or a different intervention policy,
and the same written work can be shown to tutors with and without visual
access. The dataset links evolving handwriting, tutoring conversations,
paper analyses, proactive messages, and learning outcomes for 117
participants, across 699 problem trajectories and 255,984 timestamped
images. Existing resources provide structured mathematical
dialogue~\cite{macina2023mathdial}, static images of student
work~\cite{baral2025drawedumath}, visually grounded geometry
tutoring~\cite{chowdhury2026geodial}, and multimodal tutoring
evaluation~\cite{yang2026mmtutorbench}. Ours adds real learners'
evolving work on paper with requested and proactive assistance, task
state, and test performance. The open questions it supports include when
a tutor should intervene, whether learners take up a correction, and how
well a model reads real handwriting. 
The de-identified dataset is available in the project repository (Appendix~\ref{app:tutor-materials}), together with a sample of 10 participants viewable in a dashboard (Appendix~\ref{app:trajectory-dashboard}).

\section{Limitations and Future Work}
\label{sec:limitations}

Participants were US adults on Prolific who had studied calculus before,
working on their own devices in their own spaces. The findings may not
transfer directly to students in classrooms, to first-time learners, or
to people with different access needs. The sample describes people who established a readable paper view, so the setup exits qualify every result and not only the setup-ease ratings. Every condition required
the camera, so the study does not estimate its burden relative to plain
chat tutoring. We tested one model with one set of tutoring and initiative policies; better ones are surely possible. The proactive policy as delivered was almost entirely error repair, and its trigger was not varied. Transition support was rarely triggered, and we did not analyze how often its persistence condition was met, so it is not evaluated here.

The 35-minute activity and immediate tests capture a first encounter
with the tutor. Because every condition included tutoring, the gains
include the effect of practice and cannot isolate the tutor's
contribution. Roughly 330 participants would be needed for 80\% power
to detect a prior-knowledge interaction of the size we observed.
Help-seeking purposes and tutor moves were labeled by a model
validated against three human annotators, so not every message was
labeled by humans. The thematic analysis was carried out by one researcher. The experience measures are self-reports given
immediately after the session by learners who had been told before the
activity what their tutor could do, which may shape the ratings.
Whether learners acted on each tutor message was not measured; the recorded paper trajectories make that uptake analysis possible, and it is our next step. Repeated sessions and delayed tests would then show whether the effects of a first encounter persist. A larger sample
spanning a wider range of prior knowledge could test whether proactive repair helps the learners least able to notice their own
errors.

\section{Conclusion}

We contribute a platform in which a phone camera gives an AI tutor a
live view of learners' work on ordinary paper, and a dataset that aligns evolving
handwriting with tutoring conversations and learning outcomes. In a randomized study with 117 adults, we compared three tutors: chat alone, chat plus the paper view, and the paper view plus permission to initiate help. Visual access shifted the tutor's messages toward correctness feedback on existing work, and learners with the view rated the help as more useful because they did not need to restate their steps. Initiative added error identification and directive hints, and recipients rated the proactive help supportive. Learners in every condition improved significantly from pre-test to post-test, and we did not detect a significant difference in adjusted post-test scores between conditions; an exploratory trend favored proactive tutoring for learners with low prior knowledge. These findings inform when a tutor should act on what it sees and what it should address, and the released platform and dataset support further research on the next generation of such tutors.
\FloatBarrier
\bibliographystyle{ACM-Reference-Format}
\bibliography{references}

\appendix
\section{Platform Development}
\label{app:tutor-policy}

\subsection{Guided Camera Setup}
\label{app:setup-screens}

\begin{figure*}[!t]
\centering
\includegraphics[width=.75\linewidth,alt={Two screens from the guided camera setup. The first explains how to prepare the workspace and support the phone above the paper, with illustrations. The second provides a QR code and link for opening the camera interface in the phone's browser.}]{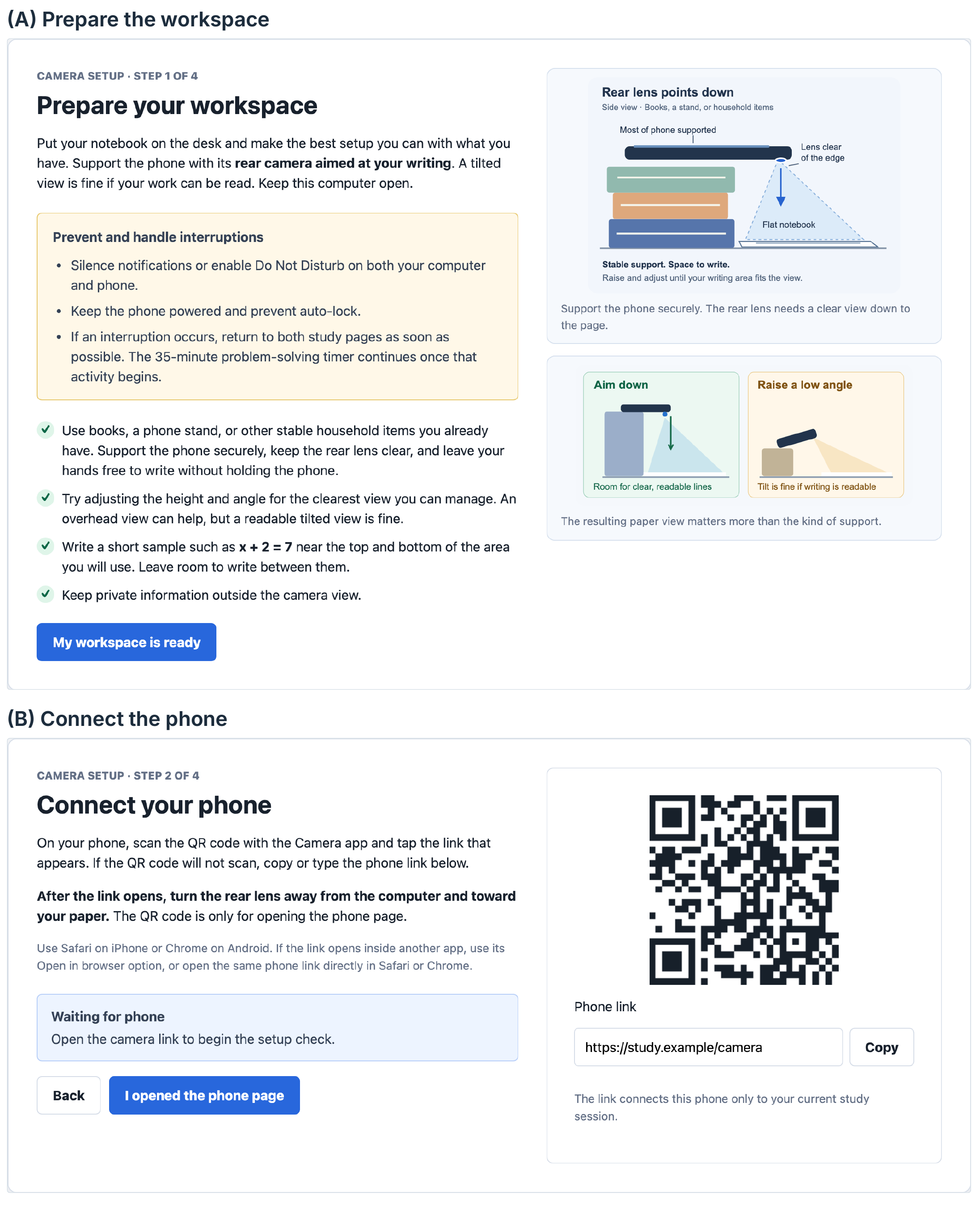}
\caption{Selected screens from the guided camera setup: (A) preparing the
workspace and (B) connecting the phone through a QR code. Screens were
rendered from the application with illustrative session details.}
\Description{Two laptop screenshots show workspace instructions and
phone-support illustrations, followed by a QR code and phone link.}
\label{fig:setup-screens}
\end{figure*}

The laptop guides learners through preparing the workspace, connecting
the phone, starting its camera, and checking the paper view.
Illustrations show how household supports can position the phone;
a QR code opens the camera page; and a laptop preview lets learners
check readability while writing. Figure~\ref{fig:setup-screens} shows two of the four setup screens, rendered with illustrative session details.

\subsection{Tutoring Prompts and System Code}
\label{app:tutor-materials}

The project repository, \url{https://github.com/nachiketdk/proactive-ai-tutoring}, contains the materials described here and in Appendices~\ref{app:instruments} and~\ref{app:trajectory-dashboard}, together with the dataset described in Section~\ref{sec:dataset}; its \texttt{README.md} maps the folders to the paper and explains their use. The \texttt{study-materials/} folder holds the setup and tutorial instructions, both test forms with scoring keys, the assisted problems with worked solutions, and the background and final questionnaires (Appendix~\ref{app:instruments}), and the \texttt{trajectory-sample/} folder holds the dashboard sample (Appendix~\ref{app:trajectory-dashboard}). The \texttt{tutoring-prompts/} folder provides the shared pedagogical policy,
the requested-help prompts for \CBase{} and \CSee{}, and the proactive
prompt for \CAct{}. The \texttt{system-code/} folder provides the tutoring modules
and participant-interface source, including problem and paper context,
proactive decisions, and worked-solution handling. A local demo
exercises these components through a provider-neutral model adapter;
recruitment, payment, and deployment services are omitted. The README
distinguishes the demo interface from the original participant client.

\paragraph{Proactive decision parameters.}
In \CAct{}, the server scheduled a paper analysis on every seventh
received paper view, about once every seven seconds at the roughly
one-view-per-second capture rate, with at most 300 analyses per session.
Each analysis supplied the latest view, earlier views for comparison, the
conversation and proactive-message history, and a computed flag indicating
whether the same work had been observed for the 120-second waiting
interval. No analysis started and no proactive message was delivered within 45 seconds of any learner or tutor message, and a proactive message prepared before the learner sent a new message was withheld. Model reasoning effort was
set to medium. These values were set during pilot rehearsals rather than
derived from prior literature.

\paragraph{Illustrative applications of the tutoring policy.}
A learner might model a rectangle's fixed area as $lw=12$, where $l$ and
$w$ are its changing length and width, then write $l'w'=0$ (primes denote
time derivatives) and ask whether the differentiation is correct. In
\CBase{}, the tutor can check the expression if it is shared in chat.
Visual access allows it to locate the product-rule error on paper even
when the expression has not been typed, and explain why both changing
dimensions matter. In \CAct{}, the visible error can prompt help before
the learner asks.

Proactive help can also address an unfinished solution without a visible
error. A learner might draw a correct diagram but not yet use its geometric
relationship to express the volume; a hint can help connect these steps.
These examples illustrate the instructed policy; the complete prompts
specify its assistance and waiting rules.

\subsection{Pilot-Informed System Refinement}
\label{app:platform-refinement}

Pilot sessions prompted two refinements. Participants arranged their
phones and paper in different ways, so handwriting was sometimes rotated
in the paper view, which led to misreadings by the tutor; we added
automatic orientation correction before views reached the tutor. We also
reduced the delay between observing the learner's work and delivering
help. We checked both changes by replaying pilot sessions and running
end-to-end test sessions, examining mathematical correctness, use of the
available context, adherence to the tutoring policy, and the delivery of
proactive messages.

\section{Test and Survey Instruments}
\label{app:instruments}

\subsection{Eligibility and Background Questions}

Eligibility required prior study of differentiation, through completed
or current calculus coursework or structured self-study, and a correct
answer to at least one of two basic differentiation questions.

The background questionnaire asked participants for their preferred
medium for working on mathematics, when they last worked with
derivatives, how often they had used AI for mathematics in the past
year, and which mathematics courses they had completed. The
\texttt{study-materials/} folder of the project repository gives the full wording and
response options; Table~\ref{tab:participant-background} states how response options were combined for presentation.

\subsection{Assisted Problem Sequence}
\label{app:curriculum-sequence}

Table~\ref{tab:curriculum-sequence} describes the conceptual opportunities
in the common problem sequence. The first four problems address the concepts the tests assess; problems five to ten provide further practice and more integrated applications. These are opportunities provided by the
materials; a correct numeric answer does not establish mastery of every
component, and participants progressed through different numbers of
problems during the fixed activity period.

\begin{table}[!htbp]
\centering
\small
\caption{The common assisted problem sequence and its conceptual focus.}
\label{tab:curriculum-sequence}
\begin{tabularx}{\linewidth}{r l >{\raggedright\arraybackslash}X}
\toprule
Order & Problem & Conceptual focus \\
\midrule
1 & Hemispherical salt mound & Distinguishing volume scaling from volume-rate scaling. \\
2 & Changing rectangle & Constructing a fixed-area relationship and determining dimensions and rates at a given instant. \\
3 & Filling a cone-shaped tank & Using similar geometry and converting a volume rate into a depth rate. \\
4 & Checking a rate comparison & Revising a proportionality inference when a derivative contains a constant term. \\
5 & Boat and rope & Relating changing lengths through a geometric constraint; distinguishing signed rate from speed. \\
6 & Power with a student error & Applying the product rule to two changing factors and recovering a missing value. \\
7 & Trapezoidal trough & Relating depth to a changing water-surface width and volume. \\
8 & Two cars & Combining signed contributions to the separation rate on perpendicular roads. \\
9 & Wedge pool with gauge & Modeling partial filling and composing rates through a sensor relationship. \\
10 & Spherical tank with gauge & Combining nonlinear volume, depth, and sensor relationships. \\
\bottomrule
\end{tabularx}
\end{table}

\subsection{Test Forms and Final Survey}

Each nine-item test form contained five multiple-choice items, three
numeric-response items, and one two-part multiple-choice item. Pilot responses and completion times informed the final item wording, scoring, and time limits.

Every survey item offered a ``Not applicable / Could not judge'' option. Items about the interpretation of written work were administered only in the visual conditions, and items about proactive guidance only in \CAct{}. Participants could record or type their open responses; recordings were transcribed for analysis.

\subsection{Test Reliability}

We calculated KR-20 from the nine binary item scores on each test form,
with responses pooled across conditions within each administration
(Table~\ref{tab:test-reliability}). We retained the scoring used in the
learning-gain analyses, including zero credit for unsubmitted and
``I don't know'' responses. To quantify uncertainty, we calculated 95\%
percentile bootstrap confidence intervals from 20{,}000 participant
resamples within each form and administration~\cite{kelley2016confidence}.

\begin{table}[!htbp]
\centering\small
\caption{Internal consistency of the nine-item test scores by form and administration. Each participant completed one form at pre-test and the other at post-test; $n$ is the number completing that form at that administration. Intervals are 95\% percentile bootstrap confidence intervals.}
\label{tab:test-reliability}
\begin{tabular}{llrrr}
\toprule
Administration & Form & $n$ & KR-20 & 95\% CI \\
\midrule
Pre-test & A & 57 & .806 & [.706, .866] \\
Pre-test & B & 60 & .846 & [.780, .891] \\
Post-test & A & 60 & .840 & [.772, .888] \\
Post-test & B & 57 & .766 & [.633, .843] \\
\bottomrule
\end{tabular}
\end{table}

\section{Interaction Coding Framework}
\label{app:process-specification}
\subsection{Coding Unit and Applicability}

Each tutor message formed one coding unit within a problem trajectory (Section~\ref{sec:process-method}). Help-seeking purposes described
the initiating learner message; tutor moves described the tutor's message. Proactive messages had no initiating
learner request. Coding used the task, reference solution, preceding
conversation, and relevant paper views.

We first identified whether the learner's message requested help.
Purpose labels applied only to help requests; tutor moves applied to
every tutor message. Each applicable purpose and move was
coded as present, absent, or cannot determine, so a message could carry
several purposes or moves. \emph{Other} indicated an identifiable function outside the listed categories; \emph{cannot determine} indicated insufficient evidence.

\subsection{Codebook}

The categories adapt distinctions from tutoring questions and feedback
research~\cite{graesser1994question,nye2014autotutor,shute2008formative},
retaining overlapping functions as in dialogue-act
coding~\cite{chen2011exploring}. Tables~\ref{tab:process-purpose-codebook}
and~\ref{tab:process-move-codebook} define the six help-seeking purposes
and eight tutor moves.

\begin{table}[!t]
\caption{Help-seeking purposes. A request can carry more than one purpose.}
\label{tab:process-purpose-codebook}
\small
\begin{tabularx}{\linewidth}{@{}>{\raggedright\arraybackslash}p{.28\linewidth}>{\raggedright\arraybackslash}X@{}}
\toprule
Purpose & Definition \\
\midrule
Verification & Requests a check of an answer, expression, approach, or interpretation. \\
Procedural guidance & Requests a formula, method, next step, or help carrying out a mathematical operation. \\
Explanation & Requests mathematical meaning or a reason for a result or operation. \\
Error diagnosis & Asks where or why existing work may be wrong. \\
Task clarification & Asks about task wording, givens, or the quantity, comparison, time, or units requested. \\
Other & Requests assistance outside these purposes, such as submission help or coordinating the camera view. \\
\bottomrule
\end{tabularx}
\end{table}

\begin{table}[!t]
\caption{Tutor moves. Labels describe the function attempted.}
\label{tab:process-move-codebook}
\small
\begin{tabularx}{\linewidth}{@{}>{\raggedright\arraybackslash}p{.28\linewidth}>{\raggedright\arraybackslash}X@{}}
\toprule
Move & Definition \\
\midrule
Correctness feedback & Affirms or rejects existing learner mathematics or a proposed interpretation. \\
Error identification & Locates or describes a specific mistake the tutor asserts is present in the learner's work. \\
Directive hint & Directs attention or suggests an operation while leaving that operation for the learner to perform. \\
Elicitation & Invites the learner to communicate work, an answer, or reasoning to the tutor. \\
Direct mathematical provision & Supplies a formula, procedure, relationship, corrected expression, or worked step. \\
Explanation & Supplies a mathematical reason or interpretation beyond the action or answer itself. \\
Task clarification & Interprets task wording, givens, or the requested quantity, comparison, time, or units. \\
Other & Provides assistance outside these moves, such as interface help or conversational coordination. \\
\bottomrule
\end{tabularx}
\end{table}

Labels capture distinct functions that may co-occur (Section~\ref{sec:process-method}). A directive hint leaves an operation to the
learner; supplying
its result is direct mathematical provision. Provision also includes
formula reminders and therefore does not necessarily mean completing
the learner's solution.

\section{Annotation Evaluation}
\label{app:calibration-development}
\subsection{Coding and Evaluation Samples}

Three annotators refined the codebook on 39 tutor messages from 16
problem trajectories and then independently coded an evaluation set of
155 messages from 32 further trajectories, one per participant, stratified by condition and problem position. The evaluation set contained 11
\CBase{}, 11 \CSee{}, and 10 \CAct{} trajectories, comprising 143 requested replies and 12 proactive messages. Annotators used the codebook, the
conversation, the task and reference solution, and the recorded paper
views. Each annotator's judgments were retained and compared separately
with the annotation model.

\subsection{Annotation Model}

The annotation model was GPT-5.6 Terra with medium reasoning. A first
pass transcribed the recorded handwriting, marking uncertain readings; a
second pass received the codebook, the task and reference solution, the
preceding conversation, the focal messages, the relevant paper views,
and the transcription, and produced the labels. Prompts and the context
procedure were developed on the 39-message set and frozen before the
155-message evaluation~\cite{pangakis2025keeping}. Inputs excluded paper views recorded after the tutor's message, later conversation and
outcomes, condition labels, and human annotations.

\subsection{Agreement}

Agreement used Krippendorff's
$\alpha$~\cite{hayes2007standard,artstein2008agreement}, with the MASI
distance for purpose and move label
sets~\cite{passonneau2006masi,ferschke2012behind,heddaya2023language}
and nominal distance for individual labels. Purpose agreement used the
110 messages with an applicable help request (109 in each model--annotator pair); move agreement used all 155 messages.
Table~\ref{tab:process-label-agreement} reports agreement for each
label. Exact agreement between the model and Annotators 1--3 was
92.7\%, 94.5\%, and 90.8\% for purpose sets and 61.9\%, 56.1\%, and
56.8\% for move sets.
Agreement also varied by condition. Among the 60 \CAct{} messages in the
evaluation set, model--annotator agreement on directive hint (.57--.64)
and elicitation (.65--.85) was lower than in the reactive conditions
(.88--1.00), while correctness feedback (.86--.93) and error
identification (.79--.86) remained close to the pooled values.

\begin{table*}[htbp]
\caption{Individual-label agreement on the evaluation set. All coefficients use nominal distance.}
\label{tab:process-label-agreement}
\small
\centering
\begin{tabular}{@{}lrrrrrr@{}}
\toprule
& \multicolumn{2}{c}{Messages} & Among & \multicolumn{3}{c}{Model versus} \\ \cmidrule(lr){2-3}\cmidrule(l){5-7} Label & Human & Paired & annotators & A1 & A2 & A3 \\
\midrule
Help-request identification & 143 & 143 & 1.000 & .941 & .941 & .941 \\
\midrule\multicolumn{7}{l}{\textit{Help-seeking purposes}} \\
Verification & 110 & 109 & .987 & .961 & .961 & .981 \\
Procedural guidance & 110 & 109 & .923 & .908 & .908 & .858 \\
Explanation & 110 & 109 & .841 & .756 & .905 & .714 \\
Error diagnosis & 110 & 109 & 1.000 & 1.000 & 1.000 & 1.000 \\
Task clarification & 110 & 109 & .876 & .905 & 1.000 & .824 \\
Other & 110 & 109 & 1.000 & 1.000 & 1.000 & 1.000 \\
\midrule\multicolumn{7}{l}{\textit{Tutor moves}} \\
Correctness feedback & 155 & 155 & .966 & .961 & .936 & .949 \\
Error identification & 155 & 155 & .893 & .822 & .810 & .868 \\
Directive hint & 155 & 155 & .922 & .766 & .780 & .818 \\
Elicitation & 155 & 155 & .973 & .922 & .922 & .962 \\
Direct mathematical provision & 155 & 155 & .895 & .829 & .778 & .763 \\
Explanation & 155 & 155 & .777 & .728 & .717 & .743 \\
Task clarification & 155 & 155 & .800 & .593 & .567 & .466 \\
Other & 155 & 155 & .827 & .744 & .588 & .658 \\
\bottomrule
\end{tabular}
\par\smallskip
\begin{minipage}{\linewidth}\footnotesize A1--A3 denote the three annotators. Human $N$ requires at least two applicable judgments; paired $N$ requires applicability for the model and the specified annotator. Purposes are conditional on the help-request identification; no-help and proactive messages are not negative purpose examples.\end{minipage}
\end{table*}

\subsection{Mathematical Correctness of Tutor Messages}
\label{app:math-correctness}

The three annotators also rated the mathematical correctness of each
tutor message in the 155-message evaluation set, judged against the
task, the reference solution, and the learner's written work. Each message was rated correct, mixed (containing any incorrect mathematics), or cannot determine. Nine messages
made no mathematical claim, such as acknowledgments, and were excluded.
Of the remaining 146 messages, each annotator rated 141 or 142 as
correct (96.6--97.3\%), 0 to 2 as mixed, and 2 to 5 as cannot determine; 139 were rated correct by all three.

\section{Supplementary Statistical Tables}
\label{app:supplementary-stats}

\subsection{Exploratory Prior-Knowledge Analysis: Median-Split Check}
\label{app:median-split}
As an exploratory check on the continuous model of Section~\ref{sec:learning-results}, we dichotomized pre-test score
at the median, as in prior work comparing learning support by prior
knowledge~\cite{naik2025tailored}. The median across
all 117 participants was three of nine, giving lower (0--3, $n=71$) and
higher (4--9, $n=46$) groups; the groups are unequal because the score is
discrete. Table~\ref{tab:median-split} reports that model's pairwise contrasts. Varying the cut point from one to five did not produce a robust result: the lower-group contrast between \CAct{} and \CSee{} reached $p<.05$ at two of five cut points (three and five) and at none after Holm correction or with robust standard errors.

\begin{table*}[!htbp]
\centering\small
\setlength{\tabcolsep}{3.5pt}
\caption{Exploratory median-split check: the ANCOVA of Table~\ref{tab:learning} with a pre-test group (median split at three of nine) and its group-by-condition interaction added. Differences are adjusted post-test points with 95\% confidence intervals; $p$ is unadjusted, Holm $p$ is corrected across the six contrasts, and Robust $p$ uses robust standard errors.}
\label{tab:median-split}
\begin{tabularx}{\linewidth}{Xlrrr}
\toprule
Contrast & Difference [95\% CI] & $p$ & Holm $p$ & Robust $p$ \\
\midrule
\multicolumn{5}{@{}l}{\textit{Lower pre-test scores (0--3)}} \\
\CAct{} vs.\ \CBase{} & 0.74 [$-$0.28, 1.75] & $.156$ & $.778$ & $.209$ \\
\CAct{} vs.\ \CSee{} & 1.16 [0.04, 2.28] & $.043$ & $.259$ & $.082$ \\
\CSee{} vs.\ \CBase{} & $-$0.42 [$-$1.52, 0.67] & $.446$ & $.900$ & $.432$ \\
\addlinespace[3pt]
\multicolumn{5}{@{}l}{\textit{Higher pre-test scores (4--9)}} \\
\CAct{} vs.\ \CBase{} & $-$0.73 [$-$2.12, 0.65] & $.298$ & $.900$ & $.200$ \\
\CAct{} vs.\ \CSee{} & 0.09 [$-$1.19, 1.37] & $.893$ & $.900$ & $.879$ \\
\CSee{} vs.\ \CBase{} & $-$0.82 [$-$2.15, 0.51] & $.225$ & $.900$ & $.229$ \\
\midrule
\multicolumn{5}{@{}l}{\footnotesize Overall tests (OLS): condition $F(2,110)=1.53$, $p=.222$, partial $\eta^2=.027$; group by condition $F(2,110)=1.53$, $p=.221$, partial $\eta^2=.027$.} \\
\bottomrule
\end{tabularx}
\end{table*}

\subsection{Tests of Label Shares}
\label{app:process-tests}
Table~\ref{tab:process-tests} gives the Kruskal--Wallis tests behind the
asterisks in Figure~\ref{fig:process-labels}, with $\epsilon^2$ as the
effect size and Holm correction applied across the labels of each label set. Table~\ref{tab:process-decomposition} separates \CAct{}'s
move shares into requested replies and proactive messages and repeats the
condition test with \CAct{} restricted to requested replies.

\begin{table*}[!htbp]
\centering\small
\setlength{\tabcolsep}{3.5pt}
\caption{Kruskal--Wallis tests of label shares across conditions. Shares are means of participant shares (Section~\ref{sec:process-method}). Holm correction is applied across the labels of each label set. Pairwise comparisons are Holm-corrected Mann--Whitney $U$ tests with rank-biserial $r$, listed only where the label differs across conditions after correction; the inequality gives the direction.}
\label{tab:process-tests}
\begin{tabularx}{\linewidth}{lrrrr>{\raggedright\arraybackslash}X}
\toprule
Label & $H(2)$ & $\epsilon^2$ & $p$ & Holm $p$ & Pairwise ($r$, Holm $p$) \\
\midrule
\multicolumn{6}{l}{\textit{Help-seeking purposes}} \\
Verification & 6.5 & .061 & $.038$ & $.230$ &  \\
Procedural guidance & 4.1 & .038 & $.131$ & $.654$ &  \\
Explanation & 0.4 & .003 & $.830$ & $>.999$ &  \\
Task clarification & 1.3 & .012 & $.522$ & $>.999$ &  \\
Error diagnosis & 0.6 & .006 & $.733$ & $>.999$ &  \\
Other & 2.8 & .026 & $.244$ & $.976$ &  \\
\addlinespace[3pt]
\multicolumn{6}{l}{\textit{Tutor moves}} \\
Correctness feedback & 27.6 & .244 & $<.001$ & $<.001$ & \CSee{} > \CBase{} ($r=.53$, $p<.001$); \CAct{} > \CBase{} ($r=.63$, $p<.001$) \\
Error identification & 18.0 & .159 & $<.001$ & $<.001$ & \CAct{} > \CBase{} ($r=.50$, $p<.001$); \CAct{} > \CSee{} ($r=.45$, $p=.002$) \\
Directive hint & 16.2 & .143 & $<.001$ & $.002$ & \CAct{} > \CBase{} ($r=.45$, $p=.002$); \CAct{} > \CSee{} ($r=.47$, $p=.002$) \\
Elicitation & 15.1 & .133 & $<.001$ & $.003$ & \CSee{} < \CBase{} ($r=.41$, $p=.003$); \CAct{} < \CBase{} ($r=.38$, $p=.004$) \\
Direct mathematical provision & 4.4 & .039 & $.108$ & $.434$ &  \\
Explanation & 2.2 & .020 & $.327$ & $.982$ &  \\
Task clarification & 0.4 & .003 & $.827$ & $>.999$ &  \\
Other & 0.5 & .005 & $.774$ & $>.999$ &  \\
\bottomrule
\end{tabularx}
\end{table*}

\begin{table*}[!htbp]
\centering\small
\setlength{\tabcolsep}{3.5pt}
\caption{Tutor-move shares in \CAct{} computed over all tutor messages, over requested replies only, and over proactive messages only, with the reactive conditions for comparison (TR, VR, and VP denote \CBase{}, \CSee{}, and \CAct{}). Entries are means of participant shares (percent). The last columns test the three conditions with \CAct{} restricted to requested replies (Kruskal--Wallis, Holm-corrected across the eight moves).}
\label{tab:process-decomposition}
\footnotesize\setlength{\tabcolsep}{3pt}
\begin{tabularx}{\linewidth}{Xrrrrrrr}
\toprule
Move & TR & VR & VP all & VP requested & VP proactive & $H(2)$ & Holm $p$ \\
\midrule
Correctness feedback & 40.1 & 62.7 & 69.3 & 58.8 & 90.9 & 18.1 & $<.001$ \\
Error identification & 20.3 & 21.6 & 40.7 & 17.9 & 85.5 & 3.0 & $>.999$ \\
Directive hint & 38.6 & 34.9 & 54.4 & 38.6 & 77.8 & 0.3 & $>.999$ \\
Elicitation & 8.3 & 2.7 & 2.5 & 3.2 & 0.0 & 13.4 & $.009$ \\
Direct mathematical provision & 67.1 & 62.4 & 58.7 & 61.5 & 53.2 & 1.3 & $>.999$ \\
Explanation & 64.7 & 56.7 & 60.9 & 57.4 & 63.5 & 3.4 & $>.999$ \\
Task clarification & 19.9 & 19.4 & 20.1 & 21.6 & 29.0 & 0.3 & $>.999$ \\
Other & 1.9 & 2.8 & 1.3 & 2.1 & 0.0 & 0.8 & $>.999$ \\
\bottomrule
\end{tabularx}
\end{table*}

\section{Trajectory Dashboard}
\label{app:trajectory-dashboard}

\begin{figure*}[!t]
  \centering
  \includegraphics[width=\linewidth,alt={Three paper images paired with tutor messages from the trajectory dashboard. For T04 in Visual-Reactive, the tutor confirms differentiation but identifies substituted dimensions from the wrong time. For T07 in Visual-Proactive, the tutor identifies an incorrect derivative of the constant-area equation and prompts use of the product rule. For T10 in Visual-Proactive, the tutor distinguishes an average change in width from the requested instantaneous rate.}]{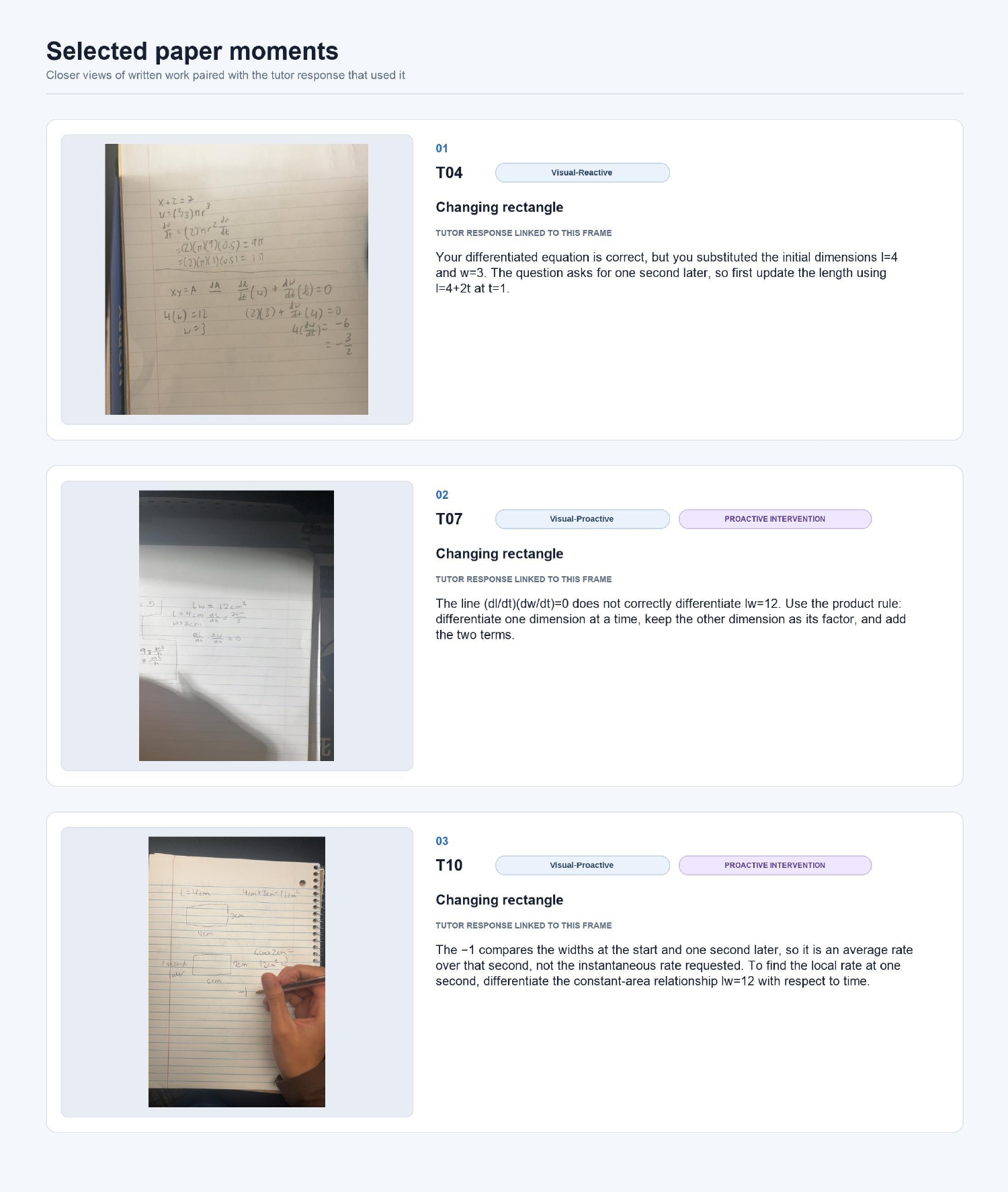}
  \caption{Three selected paper moments from the trajectory dashboard.
  Each panel pairs the paper view the tutor received with the tutor
  message that used it.}
  \Description{Three vertically stacked panels show cropped notebook images
  beside their associated tutor messages. The first is a Visual-Reactive exchange from T04. The second and third are Visual-Proactive exchanges from T07 and T10.}
  \label{fig:appendix-paper-moments}
\end{figure*}

The project repository includes an offline dashboard with
de-identified interaction examples from 10 participants spanning the
three study conditions. The dashboard connects assisted-activity
conversations to problem outcomes, pre-test and post-test responses,
final-survey ratings, and selected views of participants' written work.
The trajectories are an illustrative, de-identified sample; aggregate
analyses in the paper should be used to interpret prevalence and
condition differences. Figure~\ref{fig:appendix-paper-moments} shows
three paper moments from the dashboard at larger scale. Trajectory
identifiers such as T04 are dashboard labels unrelated to participant
numbers.

\end{document}